\documentclass{emulateapj}

\usepackage{graphicx}
\usepackage{url}
\usepackage{verbatim}
\usepackage{amsmath}
\usepackage{color,soul}
\usepackage{natbib}
\newcommand{\Lya}{Lyman-$\alpha$}

\defcitealias{Gurvich2017}{G17}
\def\paperone/{\citetalias{Gurvich2017}}

\defcitealias{Danforth:2016}{D16}
\def\DanforthCOS/{\citetalias{Danforth:2016}}

\shorttitle{Low-z \Lya~Forest \& AGN feedback}
\shortauthors{Burkhart et al.}

\graphicspath{{./}{figures/}}

\begin{document}

\title{The low redshift \Lya~Forest as a constraint for models of AGN feedback}

\author{Blakesley Burkhart}
\affiliation{Department of Physics and Astronomy, Rutgers University,  136 Frelinghuysen Rd, Piscataway, NJ 08854, USA}
\affiliation{Center for Computational Astrophysics, Flatiron Institute, 162 Fifth Avenue, New York, NY 10010, USA}

\author{Megan Tillman}
\affiliation{Department of Physics and Astronomy Rutgers University,  136 Frelinghuysen Rd, Piscataway, NJ 08854, USA}

\author{Alexander B. Gurvich}
\affiliation{Department of Physics \& Astronomy and CIERA, Northwestern University, 1800 Sherman Ave, Evanston, IL 60201, USA}

\author{Simeon Bird}
\affiliation{University of California, Riverside, 92507 CA, U.S.A.}

\author{Stephanie Tonnesen}
\affiliation{Center for Computational Astrophysics, Flatiron Institute, 162 Fifth Avenue, New York, NY 10010, USA}

\author{Greg L. Bryan}
\affiliation{Center for Computational Astrophysics, Flatiron Institute, 162 Fifth Avenue, New York, NY 10010, USA}
\affiliation{Department of Astronomy, Columbia University, 550 W 120th Street, New York, NY 10027, USA}

\author{Lars Hernquist}
\affiliation{Harvard-Smithsonian Center for Astrophysics, 60 Garden Street, Cambridge, MA 02138, USA}

\author{Rachel S. Somerville}
\affiliation{Center for Computational Astrophysics, Flatiron Institute, 162 Fifth Avenue, New York, NY 10010, USA}

\begin{abstract}

We study the sensitivity of the $z=0.1$ Lyman-$\alpha$ Forest observables, such as the column density distribution function (CDD), flux PDF, flux power spectrum, and line width distribution, to sub-grid models of active galactic nuclei (AGN) feedback using the Illustris and IllustrisTNG (TNG) cosmological simulations. The two simulations share an identical Ultraviolet Background (UVB) prescription and similar cosmological parameters, but TNG features an entirely reworked AGN feedback model. 
Due to changes in the AGN radio mode model, the original Illustris simulations have a factor of 2-3 fewer Lyman-$\alpha$ absorbers than TNG at column densities $N_{\rm HI}< 10^{15.5}$ cm$^{-2}$. We compare the simulated forest statistics to UV data from the Cosmic Origins Spectrograph (COS) and find that neither simulation can reproduce the slope of the absorber distribution. Both Illustris and TNG also produce significantly smaller line width distributions than observed in the COS data. We show that TNG is in much better agreement with the observed $z=0.1$ flux power spectrum than Illustris. We explore which statistics can disentangle the effects of AGN feedback from alternative UVB models by rescaling the UVB of Illustris  to produce a CDD match to TNG.  While this UVB rescaling is degenerate with the effect of AGN feedback on the CDD,  the amplitude and shape of the flux PDF and 1D flux power spectrum change in a way distinct from a scaling of the UVB. Our study suggests that the $z=0.1$ Lyman-$\alpha$ forest observables can be used as a diagnostic of AGN feedback models.

\end{abstract}

\keywords{}

\section{Introduction}

The \Lya~forest is a key diagnostic for the state of diffuse baryons in the intergalactic medium (IGM) and for fundamental cosmological parameters \citep{Gunn:1965,Palanque-Delabrouille:2013, McQuinn2016ARA&A..54..313M}.
    At high redshift (i.e.,  $z \geq 2$) the \Lya~forest is observed in optical wavelengths from ground-based observatories and has been used to constrain small-scale cosmic structure \citep{Croft:1998,McDonald:2005,Lidz:2010}, the IGM gas temperature \citep{Becker:2011, Boera:2014}, and the evolution of the ultraviolet ionizing background radiation \citep{Haardt:1996,Faucher-Giguere2008,Faucher-Giguere:2009, Haardt:2012,Faucher2020MNRAS.493.1614F}.
    In particular, comparison between cosmological simulations of the \Lya~forest and observations have shed light on dark matter and baryon interactions, IGM heating and cooling, and phase mixing in the IGM \citep{Cen:1994,Zhang:1995,Miralda-Escude:1996,Hernquist:1996,Rauch:1997}.
    At high redshifts ($z=2.0-5.4$), the commonly studied statistical diagnostics of the \Lya~forest such as the HI column density distribution \citep{Altay2011ApJ...737L..37A,Rahmati2013MNRAS.430.2427R}, the Doppler widths ($b$-parameter) of the \Lya~absorption lines \citep{Hiss2018ApJ...865...42H}, and the distribution and power spectrum of the transmitted flux \citep{Walther2019ApJ...872...13W,2020MNRAS.495.1825C} can be matched between simulations and observations to allow only a factor of two uncertainty in the amplitude of the metagalactic UV background (UVB).

    Early theoretical studies of the low-redshift \Lya~forest (e.g. \cite{Dave:1999}) demonstrated its potential for constraining the UVB.  However, observing the \Lya~forest at $z \leq2 $ is substantially more challenging than at higher redshifts as it requires state-of-the-art spectrographs above Earth's atmosphere to reach the rest-frame far-ultraviolet (FUV) band \citep{Danforth:2016,Gurvich2017,Viel:2017,Khaire2019,Christiansen2020MNRAS.499.2617C}.
    Recent observations with the Cosmic Origins Spectrograph (COS) aboard the Hubble Space Telescope have enabled studies of the \textit{low-redshift \Lya~forest} in great statistical detail and with high sensitivity \citep{Meiring2011ApJ...732...35M,Khaire2019,Danforth:2016,Kim10.1093/mnras/staa3844}.
    In particular, \citet[][]{Danforth:2016} (henceforth \DanforthCOS/) built on the heritage of many previous low-z IGM absorber catalogs from HST/FOS \citep{Bahcall:1996,Jannuzi:1998,Weymann:1998}, HST/GHRS \citep{Penton:2000}, FUSE \citep{Danforth:2005}, and HST/STIS \citep{Penton:2004,Lehner:2007,Tripp:2008,Danforth:2010,Tilton:2012}.
    The \DanforthCOS/ survey represents the largest sample of low-z absorbers to date, comprising 2611 distinct redshift systems and 5138 intergalactic absorption lines. In addition to the statistical significance of this survey, the exquisite sensitivity of the COS instrument has allowed for detection of very low column density absorbers in the local universe, down to column densities of $N_{\rm{HI}} \approx 10^{12.5} $ cm$^{-2}$.  

    The COS observations described in \DanforthCOS/ have challenged our theoretical understanding of the \Lya~forest in ways that have clearly called for significant further numerical study. 
    In particular, forward modeling cosmological simulations without galactic feedback or with current UVB models produces a simulated \Lya~forest that \textit{does not match} the low redshift \Lya~forest observed by COS. \cite{Kollmeier:2014} reported the first discrepancy between the observed column density distribution function (CDD) of absorbers in \DanforthCOS/ and the CDD derived using synthetic \Lya~spectra from a simulation run with the smoothed particle hydrodynamics (SPH) code GADGET. 
    \cite{Kollmeier:2014} found that the number of absorbers in simulations at low column density is too high by a normalizing factor of $3.3$ (equivalent to a factor of 5x fewer ionizing UV photons) and termed the difference the ``photon underproduction crisis.''

    \citet{Gurvich2017} (hereafter \paperone/) first suggested that feedback from active galactic nuclei (AGN) has an effect on the low-redshift \Lya~forest in cosmological simulations, finding that the simulated CDD was sensitive to the overall strength of AGN feedback, in addition to the UVB.
    When sufficiently strong AGN feedback is included, the rate of ionizing UV photons need only be increased by a factor of 1.8x rather than 5x. These photons can be accounted for in the uncertainty between different UVB models, e.g.,  the difference at z=0 between the UVB models of \citeauthor{Faucher-Giguere:2009}~\citeyear{Faucher-Giguere:2009} and \citeauthor{Haardt:2012}~\citeyear{Haardt:2012}, which differ in their assumptions of the escape fraction of UV photons from star forming galaxies. Thus, correctly modeling both the UVB and the AGN feedback in simulations is critical to matching the observed \Lya~forest.
    Later studies using different hydrodynamic simulations further confirmed that black hole feedback can heat the IGM, and is most impactful at z $< 2$ \citep{ Viel:2017,Christiansen2020MNRAS.499.2617C}, while other studies have focused on further constraining the amplitude of the UV background in the empirical models  \citep{Khaire:2015,Shull:2015,Puchwein:2018,Faucher2020MNRAS.493.1614F, Chang:2012}.

    Like the CDD, the distribution of absorption line Doppler widths (i.e. the $b$-parameter distribution) is consistently found to be mismatched between hydrodynamic simulations and observations at low redshift \citep{Viel:2017, bolton2021limits}. 
    \citet{Viel:2017} first explored the $b$ distribution for the Illustris, Sherwood, and GADGET-3 simulations and found that all three simulations \textit{underpredict} the number of absorbers at $b=25-45 $km s$^{-1}$.
    At the same time, the simulations \textit{over predict} absorbers below 20 km s$^{-1}$. 
    Using a similar analysis, \citet{bolton2021limits} found that the simulated $b$ distribution in the IllustrisTNG simulations are also systematically lower than what is observed.
 
    It is now clear that, from the perspective of numerical simulations, modeling the low-redshift \Lya~forest correctly is challenging because it requires not only: 1) a precise understanding of hydrodynamics, heating, and cooling, but also 2) accurate sub-grid models of the highly non-linear physics that governs the interactions between the gas and galactic physics (e.g. AGN feedback).
    That the statistics of the \Lya~forest at z$<2$ have been shown to be more sensitive to gas on the outskirts of collapsed objects such as filaments and galaxies further suggests that galactic physics can impact observations of the diffuse IGM \citep{Tonnesen2017,Martizzi2019MNRAS.486.3766M}. 
    In addition to the integrated strength, the \textit{nuances} of the AGN sub-grid model matter substantially for the simulated \Lya~forest properties.
    For example, the Sherwood cosmological simulations have generally found a limited impact of AGN feedback on the \Lya~forest line widths, with most hot gas from AGN too hot and highly ionized for \Lya~absorption \citep{Viel:2017, Nasir:2017,bolton2021limits}. 
    Their proposed solution is additional sources of turbulence in the IGM and heating from IGM dust.
    On the other hand, \paperone/ and \citet{Christiansen2020MNRAS.499.2617C} have found that warm, highly ionized gas produced by the jet feedback model in the SIMBA \citep{2019MNRAS.486.2827D} and Illustris simulations \citep{Springel:2018} significantly improves agreement with the mean \Lya~forest transmission at z $<$ 0.5.
    Thus \textit{the details of the sub-grid model for AGN feedback can substantially alter the phase structure of the low-redshift IGM} in cosmological simulations, even when controlling for differences in UVB models and hydrodynamic solver.

    Here, we explore the low redshift \Lya~forest in the IllustrisTNG cosmological simulation suite as compared to the original Illustris simulations in the context of the observed \DanforthCOS/ COS dataset in order to track the impact of different AGN feedback sub-grid models given the same UVB, hydrodynamic solvers, and nearly identical cosmology. 
    
    The paper is organized as follows: in Section~\ref{sec:num} we briefly describe the IllustrisTNG cosmological simulation suite, with a focus on the changes made to the AGN feedback sub-grid model compared to the original Illustris simulations.  
    In Section \ref{sec:res} we present our results comparing the statistics of the \Lya~forest in IllustrisTNG to those of Illustris and the \DanforthCOS/ data set. 
    Specifically, we compare the column density distribution, distribution of the $b$ parameter (i.e. the line widths), and flux probability distribution function and flux power spectrum.
    In Section~\ref{sec:dis} we discuss our results and summarize and conclude in Section~\ref{sec:con}.

\begin{figure*}
  
    \includegraphics[width=0.95\textwidth]{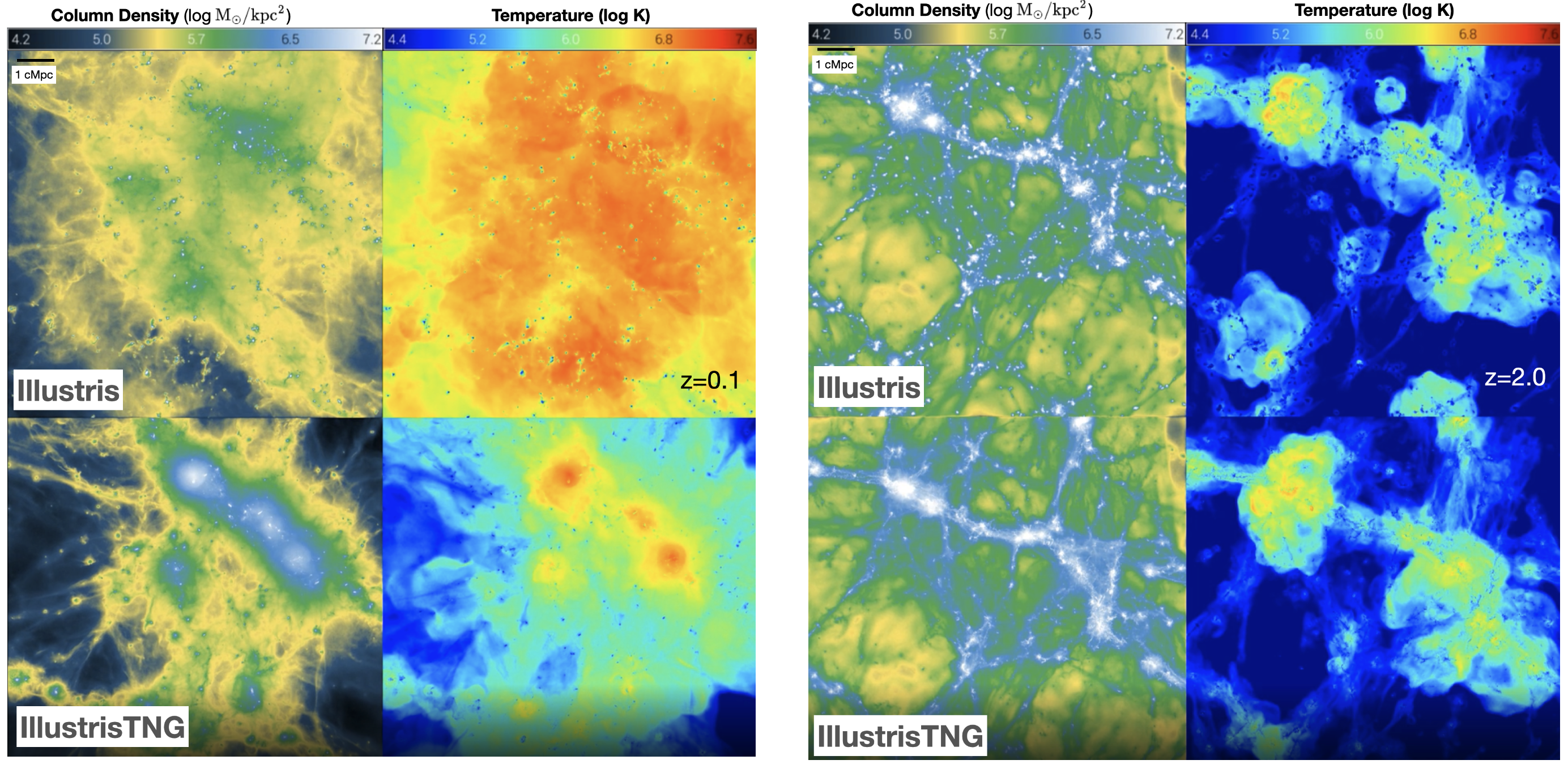}
    \caption{A visual comparison of the $z=0.1$ (left  panel) and $z=2.0$ (right  panel) of 1Mpc thick slices of the IGM column density (left columns) and  mass weighted temperature (right columns) in Illustris (top row) and TNG (bottom row).  At z=2.0, there are very minor differences between Illustris and TNG but by z=0.1 significant visual differences are obvious in the temperature and density of the IGM.
    The differences seen here are nearly entirely due to the changes in the AGN jet feedback model. 
   Hot ionized medium (HIM) gas above $10^7$ K (traced in x-rays wavelengths; red color in the temperature color-bar) dominates in Illustris while the IGM volume is more filled by warm \Lya~emitting gas in TNG (gas around $10^4$ K (blue color in the temperature color-bar).
    \label{fig:image} }
\end{figure*}

\section{Numerical Simulations} \label{sec:num}

    In this section we provide an overview of the Illustris and IllustrisTNG simulations used in this study, both of which are performed with the AREPO code \citep{Springel:2010,Weinberger2020ApJS..248...32W}. 

    Illustris is a cosmological hydrodynamic simulation in a box of length 75 h$^{-1}$Mpc.
    Gravitational interactions evolve via the TreePM algorithm \citep{Springel:2005}.
    Radiative cooling is implemented using the network described in \citet{Katz:1996} and includes line cooling, free-free emission, and inverse Compton cooling.
    Illustris assumes ionization equilibrium and accounts for on-the-fly hydrogen column density shielding from the radiation background \citep{Rahmati:2013}. Metals and metal-line cooling are included \citep{Vogelsberger:2012,Vogelsberger:2013} and star formation is implemented using the \citet{Springel:2003} subgrid model.  Illustris uses $\Omega_m $= 0.2726, $\Omega_{\Lambda} = 0.7274$, $\Omega_b= 0.0456$, $\sigma_8$ = 0.809,  and $H_0 = 100$ h km s$^{-1}$Mpc$^{-1}$ with h = 0.704. These parameters are consistent with the Wilkinson Microwave Anisotropy Probe 9 measurements \citep{2013ApJS..208...19H}.
    
    While the Illustris simulation suite has been remarkably successful in reproducing a number of observed galaxy properties \citep{Vogelsberger:2014a,Vogelsberger:2014b,Sijacki:2015}, as well as properties of high redshift neutral gas \citep{Bird:2014}, its subgrid models for black hole feedback are in tension with a number of other observed galaxy properties \citep[e.g., low gas fraction of groups of galaxies as shown in][]{Genel:2014,Nelson:2015}.
    This tension partially motivated the next generation of the Illustris simulations, IllustrisTNG (TNG), which included an entirely updated sub-grid model for AGN feedback developed in \citet{Weinberger:2017}, which we briefly describe below. 

    The TNG suite consists of 18 magnetohydrodynamic cosmological simulations which vary in mass resolution, volume, and complexity of the physics included \citep{Pillepich2018MNRAS.475..648P,Marinacci:2018,Naiman2018MNRAS.477.1206N,Springel:2018,Nelson2018,Nelson:2019a,Pillepich:2019,Nelson:2019b}. 
    Three box sizes are employed with cubic volumes of 51.7, 110.7, and 302.6 Mpc side length, referred to as TNG50, TNG100, and TNG300, respectively. TNG's default cosmological parameters are consistent with the \citet{2016A&A...594A..13P} results, with matter density $\Omega_m = \Omega_{dm} + \Omega_b = 0.3089$, baryonic density $\Omega_b = 0.0486$, cosmological constant $\Omega_{\Lambda} = 0.6911$, Hubble constant $H_0 = 100$ h km s$^{-1}$Mpc$^{-1}$ with h = 0.6774, normalisation $\sigma_8$ = 0.8159. 
    The largest box simulation (TNG300) allows for the study of rare objects/events and galaxy clustering and provides the largest galaxy sample while the smallest physical volume simulation (TNG50) allows for study of more detailed structural properties of galaxies and their kinematics as well as the opportunity to do resolution convergence testing. 
    TNG100 (also referred to in this work as simply TNG) uses the same initial conditions as the original Illustris simulation, allowing for direct comparison with the original Illustris and testing of the new physics included in the TNG runs. 
    Each of the simulation boxes has been run at three resolution levels, allowing for further convergence studies.

    The main differences between TNG and Illustris are: (I) a new implementation of stellar driven galactic winds with modified velocity, thermal content, and energy scalings \citep{Pillepich:2018b}, (II) a new implementation of black-hole-driven kinetic feedback at low accretion rates, and (III) the inclusion of magnetohydrodynamics.
    \paperone/ has already shown that galactic winds (as implemented in these models) have essentially no impact on the low-redshift \Lya~forest properties, as their effects cannot travel to substantial distances outside of halos into the IGM, so we do not expect these changes in the physics model to affect the interpretation of our results.

    Of great importance for the statistics of the \Lya~forest is the implementation of the ultraviolet background (UVB) and the related photoheating rate. 
    Illustris and TNG both use the UVB model presented in \citet{Faucher-Giguere:2009}.
Changes to the UVB model and their effects on the IGM at z$<0.5$ have been explored in other works \citep[e.g.][]{Shull:2015,Gaikwad:2017,Khaire2019,bolton2021limits} therefore we focus on studying the differences between the AGN feedback sub-grid models.
    This direct comparison is made possible by the fact that TNG keeps the simulation properties known to affect the \Lya~forest the same as they were in Illustris (e.g. the UVB model) except for the redesign of the AGN feedback model at low black hole accretion rates. 

    In the original Illustris, the AGN feedback sub-grid model is split into a two-accretion-mode scenario in which feedback energy is injected by a \textit{radio} mode at low accretion rates (relative to the Eddington limit) and a \textit{radiative} mode at high accretion rates.
    The radio mode inflates a hot explosive bubble in the circumgalactic medium whose radius scales with energy and density of the gas, however this resulted in too-low gas fractions in galaxy groups and low-mass clusters due to an over-efficient expulsion of the gas \citep{Genel:2014}.  Additionally, the mode generated too-high stellar masses in the central galaxies.
    Further details of the Illustris AGN model are described in \citet{Genel:2014} and \citet{Vogelsberger:2013}.
    By contrast, the AGN feedback model employed in TNG assumes an updated two-accretion-mode scenario in which the feedback energy is injected through (1) a relatively efficient kinetic mode at low accretion rates, and (2) a less efficient thermal mode at high accretion rates in which thermal energy cools rapidly, which is similar to that in Illustris. 
    The kinetic mode in TNG imparts a randomly oriented momentum boost (i.e. isotropic when integrated over time) to the gas cells in the few kpc feedback region around the black hole (with no mediation by wind-particles) in a series of discrete injection events, which occur once the accumulated energy output from the black hole has exceeded a threshold. 
    For more details on the TNG AGN model see \citet{Pillepich2018MNRAS.475..648P,Zinger2020MNRAS.499..768Z}.

    In Figure \ref{fig:image} (left most panels) we show a visual comparison of the z=0.1 IGM column density (left column) and temperature (right column) in Illustris (top row) and TNG (bottom row). 
    The differences seen are primarily due to the changes in the AGN jet feedback model. 
    HIM gas above $10^7$ K (traced in x-rays wavelengths; red color in the temperature color-bar) dominates in Illustris while in TNG the IGM volume has more \Lya~emitting gas (gas around  $10^4$ K, blue color in the temperature color-bar).
    In both simulations, the Warm-Hot Ionized Medium (WHIM; 10$^5$K $<$ T $<$ 10$^7$ K, n$_H < 10^{-4}$(1+z) cm$^{-3}$)  e.g. \cite{Dave:2001}) gas fills the volume many Mpcs away from the galaxies.
    However, this gas is too hot to effectively absorb in \Lya.
    Interestingly, at $z=2.0$, the IGM column density and temperature are much more similar between Illustris and TNG, as shown in Figure \ref{fig:image} (i.e., right most panels).
    As opposed to the situation at $z=0.1$, much of the volume away from the central galaxies is cold ($T\lesssim 10^4$K) and the thermal state of the IGM is controlled by the UVB.
    Comparison of these figures not only demonstrates the ability of AGN feedback to impact gas out to many Mpcs \citep[see also,][]{Gurvich2017,borrow2020MNRAS.491.6102B,Christiansen2020MNRAS.499.2617C} but specifically that the strength and/or \textit{manner in which the AGN feedback is implemented} matters substantially in determining the phase structure of the IGM. In particular the bubble model in Illustris produces copious amounts of WHIM. TNG instead has the updated kinetic AGN feedback model at low black hole accretion rates which produces less extremely hot gas in the IGM. 
    
\section{Results} \label{sec:res}
    In this section we present a comparison of the column density distribution (CDD), the distribution of the Doppler $b$-parameter, the transmitted flux probability distribution function (PDF), and power spectrum between Hubble COS spectra, TNG, and Illustris.  

    In order to generate realistic spectra from the simulations we use the publicly available code outlined in \citet{Bird:2015} and \citet{Bird:2017}\footnote{\url{https://github.com/sbird/fake_spectra}}.
    Similar to \paperone/, we estimate the CDD from the column density along 5,000 simulated sightlines randomly oriented throughout the simulation box.
    Column densities are computed by interpolating the neutral hydrogen mass in each gas element to the sightline using an SPH kernel. 
    For the column density, the sightline is split into absorbers every 50 kms$^{-1}$ (we confirm that our results are not sensitive to this value).
    In addition to this ``direct'' CDD method, we also compute Voigt profile fits to the optical depth of our simulated spectra (using a fitting algorithm included in the \texttt{fake\_spectra} package used to generate the simulated spectra) to obtain best-fit column densities and $b$ parameters. \paperone/ showed that there is no substantial difference between the Voigt and ``direct'' CDD calculations, and therefore we will omit such a comparison here.

\subsection{The Column Density Distribution}
    We define the CDD function:
    \begin{equation}
      \frac{d^2N}{\mathrm{d} \log(N) \mathrm{d}z}=\frac{F(N)}{\Delta\log( N)}\Delta z  
    \end{equation}
    F(N) is the number of absorbers per sightline with column density in the interval [N$_{HI}$;N$_{HI}$ + dN$_{HI}$], over the redshift interval $\Delta z$ contained in our simulated spectra and $\Delta\log( N)$ is the logarithmic bin width. 

\begin{figure}
    \centering
    \includegraphics[width=0.45\textwidth]{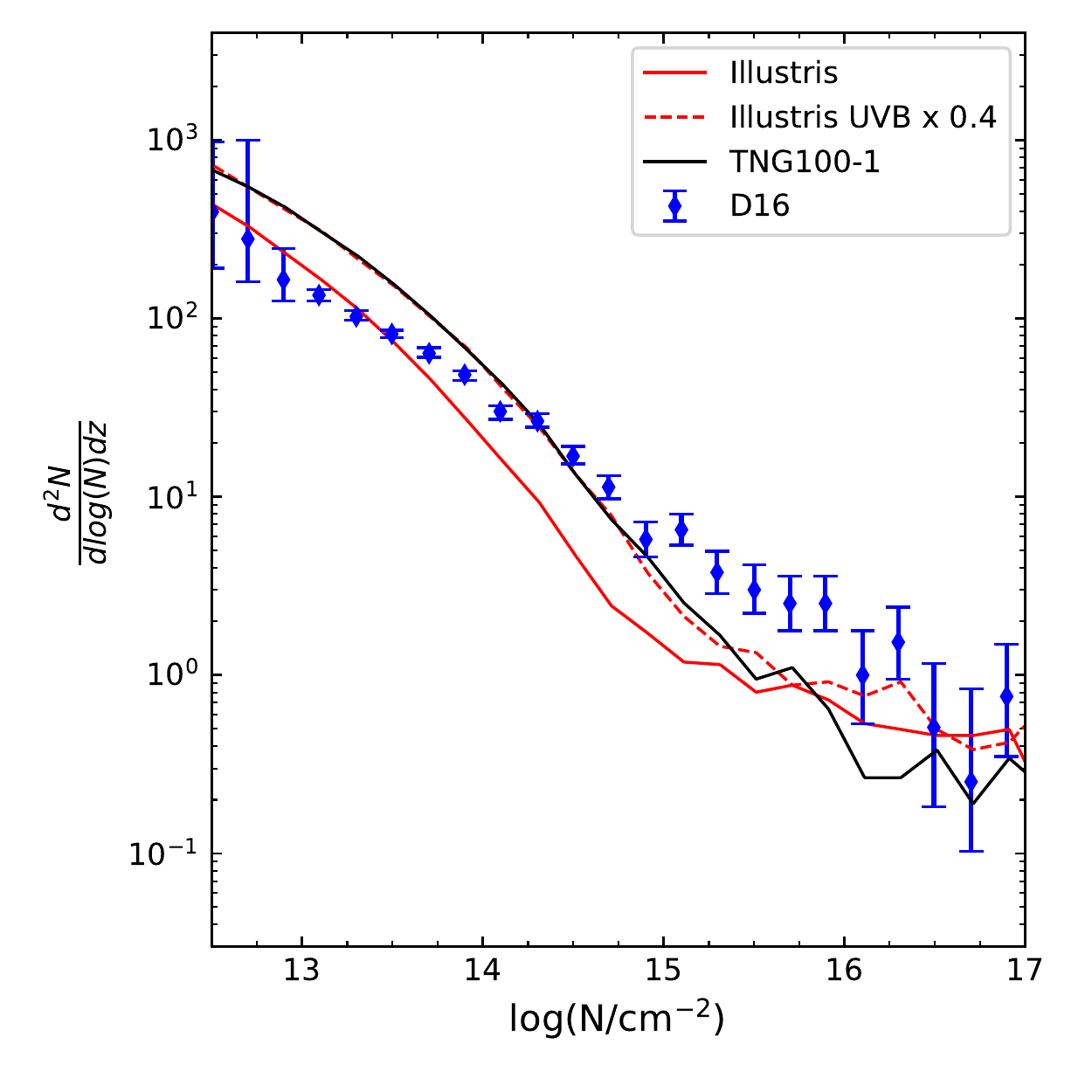}
    \caption{A comparison of the z=0.1 CDD between Illustris (red solid line) and TNG (black solid line) with the \DanforthCOS/ data set (blue points). 
    Both Illustris and TNG use the fiducial boxes. In order to see if the altered AGN feedback model could mimic the effects of the UVB, we include the CDD of Illustris with a UVB field strength $0.4$ times the original (i.e. dashed red line). Changing Illustris UVB by this factor provides a good match to TNG for column densities below $10^{15}$cm$^{-2}$}
    \label{fig:mainCDD} 
\end{figure}
    
    Figure \ref{fig:mainCDD} shows a comparison of the $z=0.1$ \Lya~forest CDD of Illustris (i.e. the results of \paperone/; solid red line), new measurements of the same from TNG (solid black line), and the \DanforthCOS/ data set (blue points). We also include Illustris modified with a UVB field strength 0.4 times the original (dashed red line).
    We find that neither Illustris nor the updated TNG simulations can reproduce the COS CDD of the low redshift \Lya~forest.
     As reported by \paperone/, we find that the Illustris CDD (solid red line) matches the \DanforthCOS/ data well in the column density range of N$_{\rm HI} = 10^{12}$-$10^{14}$cm$^{-2}$.
    TNG (solid black line) has too many absorbers in this column density range to match the observations, indicated by a higher amplitude in the CDD.
    Ultimately, we find that the CDD slope of both TNG and Illustris does not match the 
    \DanforthCOS/ data. In particular, the slope of both TNG and Illustris is too steep to match the COS data. 
    
      We find that AGN feedback can shift the amplitude of the CDD, in a similar way as when altering the UVB. We demonstrate this by modifying the Illustris \Lya~ flux with UVB field strength 0.4 times the original Illustris simulation UVB (dashed red line). This is done in post processing by recalculating the neutral hydrogen fractions using scaled UVB values. We choose the value of 0.4 in order to produce an Illustris CDD. which matches well with TNG in the column range of $\rm{log N}_{HI} < 15$. 
      There remains a slight difference in the slope of TNG and Illustris CDD at $\rm{log N}_{HI} > 15$, however in this saturated regime the determination of column densities is
more sensitive to details of the line fitting method, spectral resolution, and it is subject to greater statistical uncertainties.

    \citet{Viel:2017} investigated altering the UVB to force a match between Illustris (and other simulations such as the Sherwood simulations) and the COS data, finding that doing so requires an aggressive change in the photoionization rate of a factor of 1.5-3 times higher than \citet{Haardt:2012}. 
    However, the increase in required photons for TNG is not as extreme as models that lack AGN, e.g., the simulations used in \citet{Kollmeier:2014} required a factor of 5 more UV photons than \citet{Haardt:2012}.
    Including AGN feedback is a possible solution to requiring a severe increase in the ionization rate as it contributes to heating the gas and therefore reduces the temperature dependent recombination rate coefficient until the gas is no longer able to absorb in \Lya.
   Therefore, it is likely that the full resolution of the discrepancy between the simulated and observed CDDs will require \textit{both} an alteration in UVB and a reworked AGN feedback model that is tuned to both galaxy halo properties and properties of the IGM.

\subsection{The Doppler $b$ Parameter}

\begin{figure*}
    \centering
    \includegraphics[width=0.65\textwidth]{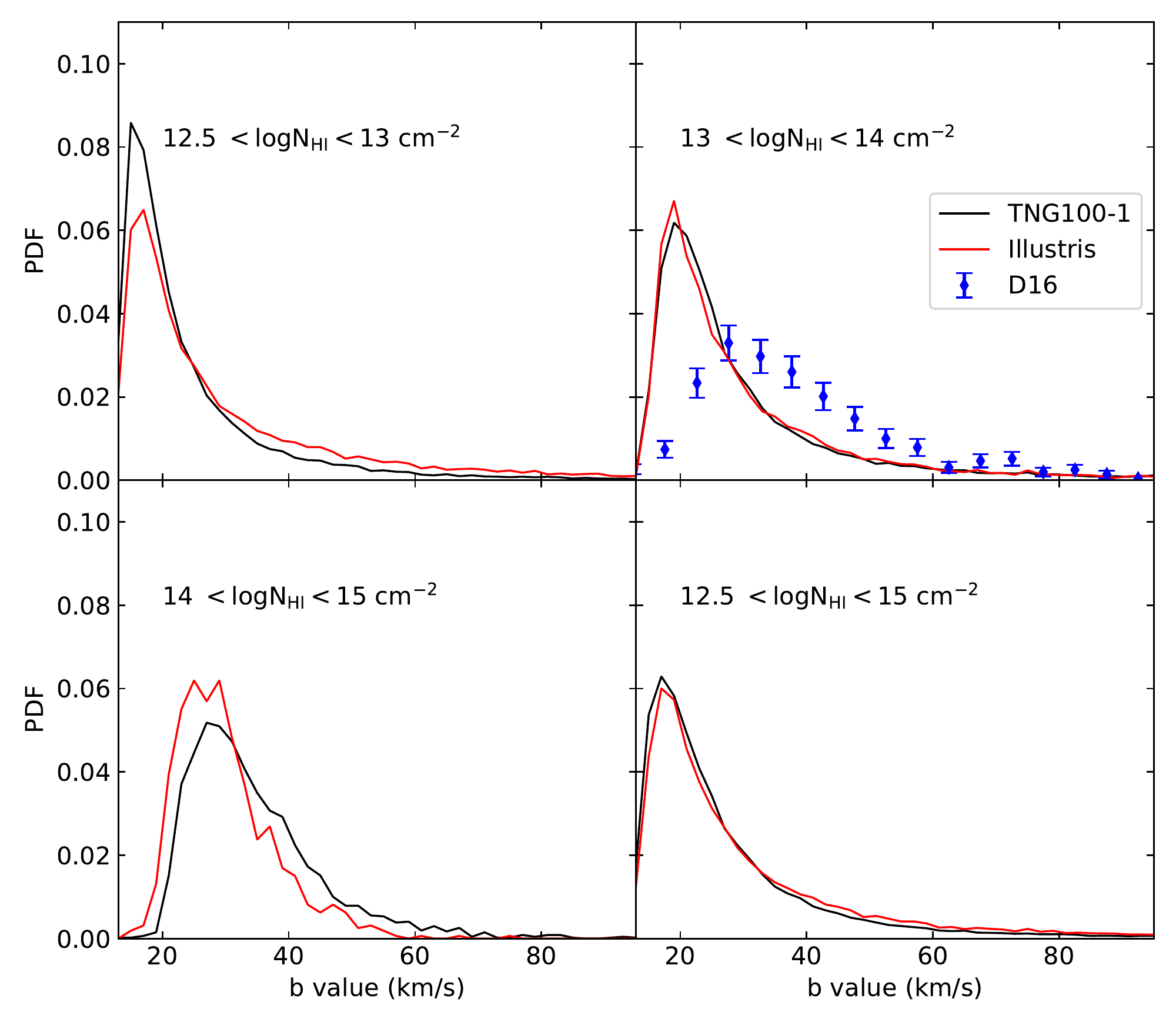}
    \caption{ Probability distribution function of the Doppler ($b$) parameter comparing TNG (black line), and Illustris (red lines).
    From top left to bottom right the panels show the distribution broken down by column density ranges. 
    The bottom right shows the full range of column density for the low and intermediate column density range of the IGM ($10^{12.5} \leq$ N$_{HI} \leq 10^{15}$ cm$^{-2}$). The top right plot includes the COS observations first analyzed in \citet{Danforth:2016} (pink points) and later reanalyzed in \citet{Viel:2017} (blue points) in the same column density bin.
    For each plot $b$ values less than 13 km/s were removed as that approximately corresponds to the cooling floor of the simulations. 
    Values higher than 100 km/s were removed as $b$ values higher than this are typically not observed and also tend to correspond to bad Voigt fits.}
    \label{fig:bTNG} 
\end{figure*}

    We show the Doppler $b$ parameter distribution for Illustris (red lines) and TNG (black lines) in Figure \ref{fig:bTNG}. 
    In the bottom right panel we show the entire range of column density values we resolve (12.5 $< \rm{log N}_{HI} < 15$ cm$^{-2}$). 
    Note that fitting Voigt profiles at column densities lower than $\rm{log N}_{HI} < 12.5$ is difficult to perform accurately, while column densities higher than $\rm{log N}_{HI} > 15$ become increasingly rare and suffer from low number statistics.
    When considered over the entire column density range, Illustris and TNG show almost identical $b$ distributions peaked at $b\approx 20$ km s$^{-1}$ with a long tail to large values.
    For reference, $b = 22$ km s$^{-1}$ is equivalent to thermally broadened lines with T=10$^{4.5}$ K.
    The similarity in the two $b$ distributions suggests that, when considered over the entire column density range, the $b$ distribution for the \Lya~ Forest is largely insensitive to the sub-grid AGN feedback model.
    
    Other panels in Figure \ref{fig:bTNG} consider low, intermediate, and high column density subsets of the full column density range to determine if there is a regime in which the effects of different sub-grid AGN feedback models can be distinguished.
    At the lowest column density range considered (top left; 12.5 $< \rm{log N}_{HI} < 13$ cm$^{-2}$) Illustris tends to have slightly higher $b$ values than TNG but both are still peaked at $20$ km s$^{-1}$.
    Higher b values in Illustris are expected because its AGN radio feedback model produces an IGM with more volume filling hot bubbles than TNG (see Figure \ref{fig:image}).
    At the opposite end of the spectrum at higher column density values (bottom left; N$_{\rm HI}= 10^{14}$-$10^{15}$cm$^{-2}$), both Illustris and TNG are peaked at $30$ km s$^{-1}$ though the $b$ distribution for Illustris is slightly less skewed than TNG.
    The skew in TNG is likely due to shock heating from the AGN kinetic mode, which can affect these higher IGM column densities (see Figure \ref{fig:image}).
    
    In the intermediate column density range (top right; N$_{\rm HI}= 10^{13}$-$10^{14}$cm$^{-2}$) Illustris and TNG are identical. 
    As a point of comparison, we also overplot the COS observations analyzed in \DanforthCOS/ (blue points) in the same column density range.
    It is clear that, as in other studies, both Illustris and TNG produce $b$ distributions much lower than what is observed \citep{Gaikwad:2017,Nasir:2017,bolton2021limits}.
    Though the simulated spectra have not been corrected to match the spectral resolution of COS (FWHM=15 km/s), the differences seen between simulation and observation in Figure \ref{fig:bTNG} are too large for this to be the sole solution.
    One possible reason for the difference in $b$ values between the simulations and observations is that the simulations are underresolving the amount of turbulence present in the IGM or are missing sources of turbulence, which we discuss further in Section \ref{sec:bturb}.
    
    Overall, though there are small differences at the lowest and highest column densities we resolve, we find that the overall $b$ distribution is not affected by differences in the sub-grid AGN feedback model employed. We note that the Illustris $b$ distribution is largely unchanged when altering the UVB.

\subsection{The IGM Density-Temperature Relation}
\begin{figure*}
\centering
\includegraphics[width=0.95\textwidth,trim={0cm 0cm 0cm 0cm},clip]{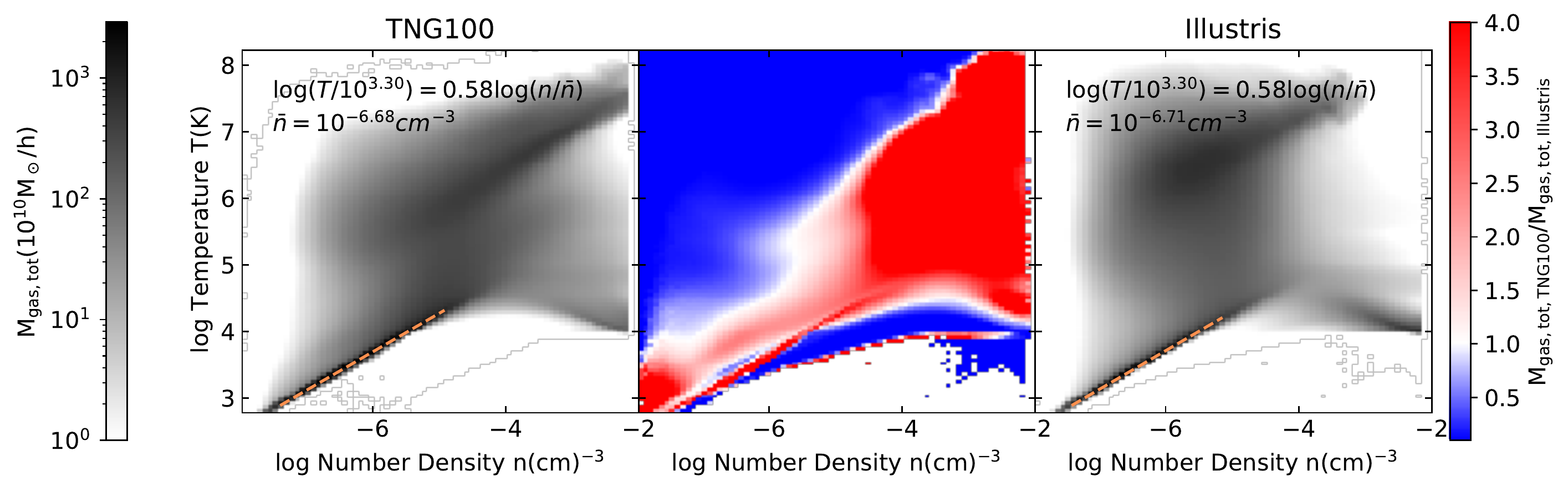}
\caption{Mass-weighted temperature-density contour plot for  hydrogen gas in the IGM in Illustris (far right) and TNG (far left) and their ratio (center). The central plot is colored by the gas mass ratio of TNG to Illustris.}
\label{phase}
\end{figure*}

    To further understand the effects of the sub-grid AGN feedback model on the IGM temperature and column density we plot the 2D temperature-density relation (otherwise known as the equation of state) for IGM gas in Figure~\ref{phase}.
    In simulations of the diffuse IGM, the temperature and density are known to satisfy a power-law relation over two decades in density, $T(\Delta)= T_0 \Delta^{\gamma-1}$ where $\Delta=\rho/\rho_0$ is the overdensity factor, $\gamma$ is the power law index, and $T_0$ is the temperature at the mean density \citep{Katz:1996,Hernquist:1996,Dave:1999}.
    This power-law relation quantifies the thermal state of the IGM \citep{1997MNRAS.292...27H,McQuinn2016ARA&A..54..313M}. 
    
    In Figure \ref{phase} we show the phase diagrams for TNG (far left), Illustris (far right), and the ratio of the two (center). 
    The ratio plot is colored by the gas mass ratio of TNG to Illustris, i.e. red identifies regions of temperature-density phase space more populated in TNG and blue those more populated in Illustris. 
    White color indicates the two simulations are the same in that part of phase space.
    The power law region of the diffuse IGM ($10^{3}~\mathrm{K} \leq T \leq 10^{4.5}$ K; $10^{-8}~\mathrm{cm}^{-3}\leq n \leq 10^{-5} ~\mathrm{cm}^{-3}$) is nearly identical between the two simulations. This should be expected as the b-value distributions are very similar between the two simulations. However looking at the middle panel of Figure \ref{phase} around the power law, the IGM gas appears to be warmer in Illustris than in TNG. 
    Given that these two simulations employ the same UVB, another source of heating is needed to explain the temperature shift near the power law relation.  The additional heating is likely due to the hot AGN bubble model in Illustris.  
    
    Looking at higher temperatures, more gas in Illustris is pushed into the HIM/WHIM phase-state as compared to TNG: \citet{Martizzi2019MNRAS.486.3766M} found that the mass of the WHIM in Illustris is ${\sim}26\%$ larger than in TNG, consistent with this excess. 
    However, this excess is concentrated at either high density (i.e. in the circumgalactic gas) or at the very lowest densities.  
    In Figure \ref{phase}, intermediate densities ($-4 \leq \log(n) \leq -7$ cm$^{-3}$) are white in color, suggesting little to no difference between TNG and Illustris here. 
    Illustris has far more HIM than TNG. 
    However,  hot gas above $10^5$K does not absorb in \Lya~ and is instead probed better by ionized gas tracers beyond the scope of this study (e.g. ionized metal lines and x-rays).
    We note an excess of T=$10^{4}$K gas at high density in Illustris that is absent in TNG. This gas is halo/ISM gas that is not relevant for the IGM and is discussed further in \citet{Martizzi2019MNRAS.486.3766M}.

\subsection{The Flux PDF and Power Spectrum}

 The amplitude of the \Lya~forest power spectrum is sensitive to the UV background and can be used to measure the HI photoionization rate $\Gamma_{HI}$. 
    Recently, the power spectrum of the low redshift \Lya~forest was estimated from the COS data of \citet{Danforth:2016} and then used to estimate the photoheating rate \citep{Khaire2019}.
    Based on the power spectrum, they found a $\Gamma_{\rm{HI}}$ that agreed with previous estimates \citep{khaireSri2019MNRAS.484.4174K,Puchwein:2018,Shull:2015}. 
    Their analysis suggested that the low-z UV background is dominated by ionizing photons emitted by quasars and does not require any significant contribution from galaxies. In this section we investigate if the flux statistics contain dependencies on the AGN feedback model.

\begin{figure}
    \centering
    \includegraphics[width=0.45\textwidth]{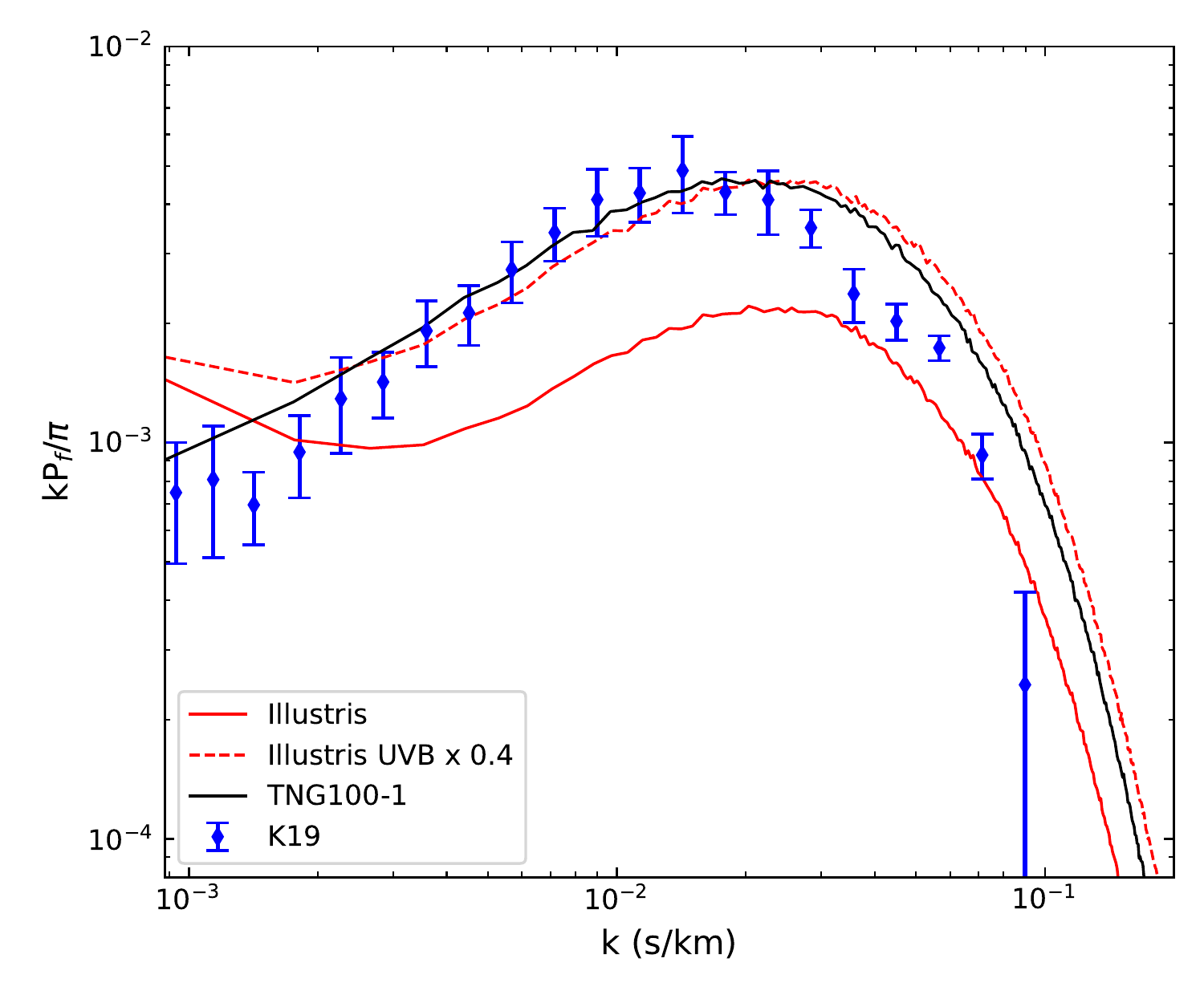}
    \caption{ Power spectrum of the z=0.1 \Lya~forest in  TNG (black lines) and Illustris (red line). The fundamental mode for the simulations is $k_{\rm{max}}=2 \pi / v_{\rm{max}} = 8.8 \times 10^{-4}$ s/km for $z = 0.1$. }
    We overplot the COS data presented in \cite{Khaire2019} in blue. The dashed red line shows Illustris with a rescaled UVB.
    \label{fig:powerspectrum} 
\end{figure}

    We present the flux power spectrum of the $z=0.1$ \Lya\ forest in TNG and Illustris in Figure \ref{fig:powerspectrum}. 
    We overplot the COS data presented in \citep{Khaire2019} in blue.
    Compared to the observations, TNG is in much better agreement than Illustris for $8.8 \times 10^{-4} < k < 0.07$ s/km. 
    Both simulations and observations show evidence of a thermal cut-off in the power at small scales ($k > 0.02$ skm$^{-1}$), which is a result of pressure smoothing of the IGM and thermal broadening of absorption lines. This cut-off is an important feature that probes the thermal state of the IGM \citep{2001ApJ...557..519Z,2019ApJ...872...13W}. The cutoff is shifted towards smaller scales in the simulations as compared to the data, reflecting the lower gas temperatures in the simulation. In particular, TNG produces more power at $k > 0.07$ s/km when compared to COS.
    TNG and Illustris are different in both shape and normalization. 
    
    The amplitude of the flux power spectrum in Illustris is significantly lower than in TNG. However, these differences can be reduced by using Illustris with a rescaled UVB selected to match the CDD (i.e. the dashed red line in Figure~\ref{fig:powerspectrum}). However, Illustris produces excess power on the largest scales probed, $k < 2\times 10^{-4}$ s/km, which could be due to large-scale correlations induced by AGN heating. From this we can conclude that the difference manifested by the altered feedback models in these simulations is not solely in the temperature distribution of the gas, but also in the fact that the physical scale of AGN feedback in Illustris is large enough to change the power spectrum significantly,  even when the mean absorption is matched. The mismatch of Illustris with the COS data (blue points) suggests that the AGN feedback in Illustris is unrealistically strong. This conclusion was also previously discussed based on a mismatch of galaxy properties \citep[e.g., low gas fraction of groups of galaxies as shown in][]{Genel:2014,Nelson:2015} and our results suggest less extreme AGN feedback models, such as TNG,  could also be tested using the \Lya~flux power spectrum.
    
    We emphasize that one \textit{cannot} attribute the amplitude and shape differences in the z=0.1 power spectrum between Illustris and TNG to the photoheating  and photoionization rates as \textit{these rates  are identical} (both simulations use the \citeauthor{Faucher-Giguere:2009} \citeyear{Faucher-Giguere:2009} UVB).
    This suggests that differences in the sub-grid AGN or stellar feedback models is responsible for the differences seen in the power spectrum. 
    We note that \paperone/ finds that the \Lya~forest is largely insensitive to stellar feedback.  Furthermore, our results suggest N-body simulations that only explore changes in photoionization rates and do not include AGN feedback processes do not capture all the subtleties that can affect the flux power spectrum and simply rescaling the power spectrum amplitude using a different UVB may not capture all the necessary IGM physics relevant at low redshift.

\begin{figure}
    \centering
    \includegraphics[width=0.45\textwidth]{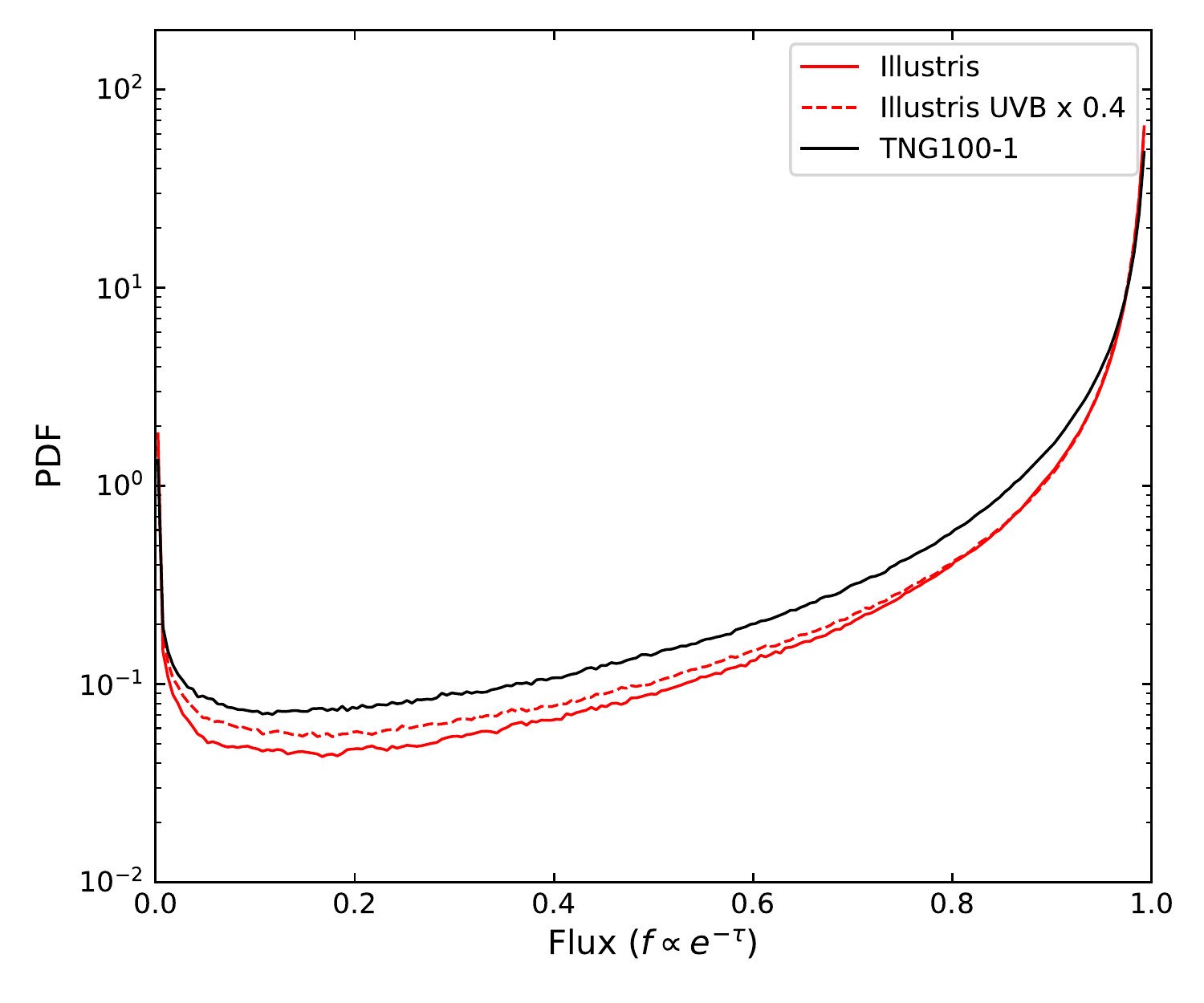}
    \caption{ Flux PDF of the z=0.1 \Lya\ forest in  TNG (black line) and Illustris (red line). The dashed red line is the UVB correction for Illustris.}
    \label{fig:fluxpdf} 
\end{figure}

    In addition to the power spectrum, we present the flux PDF of Illustris (red solid line) and TNG (black solid line) in Figure \ref{fig:fluxpdf}.  We also include the Illustris UVBx0.4 comparison which matches the CDD of TNG (dashed red line). 
    As expected from the CDD, the flux PDF amplitude in Illustris is lower than TNG until the optically thin regime.  Furthermore, we see that simply altering the UVB is not able to produce matching flux PDFs from the two simulations, as it is sensitive to absorbers with column densities too small to be Voigt fit, which are thus not included in the column density function.
    As was the case with the power spectrum, the noticeable differences between the two simulations suggest that the flux PDF may be a useful tool to constrain sub-grid models of AGN feedback.

\section{Discussion} \label{sec:dis}

    \subsection{The importance of AGN feedback on the low-z \Lya~forest}
        AGN feedback is crucial for reproducing galaxy properties \citep{Somerville2015ARA&A..53...51S}, but its effect on the physical state of the low redshift IGM is not well understood.
        The last decade has witnessed the development of various AGN feedback models for cosmological hydrodynamic simulations that are tuned to quench galaxies in agreement with observations \citep{Vogelsberger:2014, Schaye2015MNRAS.446..521S, Weinberger:2017, henden2018MNRAS.479.5385H}.
        It has only recently become clear that AGN feedback can potentially carry matter and energy far from the central regions of galaxies and have a substantial effect on the thermal state of the IGM  \citep{borrow2020MNRAS.491.6102B,Christiansen2020MNRAS.499.2617C}.
        In this paper we have demonstrated that AGN heating is impactful at low redshift and the non-linear growth of structure driving the WHIM and the diffuse IGM are subject to the details of the subgrid feedback-physics implementation. 
        TNG and Illustris employ nearly identical cosmological parameters and numerical schemes but differ dramatically in their stellar and black hole feedback prescriptions.
        The sensitivity of the low-redshift \Lya~forest to hydrodynamical effects emphasizes the importance of AGN feedback physics and related effects, which makes the low-redshift \Lya~forest an ideal test-bed for realistic hydrodynamical simulations.
        In addition to the CDD, the differences between Illustris and TNG in the flux power spectrum and PDF are significant and cannot be attributed to changes in the photoheating rate.
    
        The gas in recent numerical simulations is either too hot to contribute to Lyman-$\alpha$ absorption or too cold (in terms of temperature or amplitude of turbulence) to produce the required linewidths, fluxes and absorber distribution. 
        The radio mode (hot bubbles produced in the black hole low accretion mode) in Illustris produces an unphysically large amount of volume filling hot gas at $z<2$ and produces too few \Lya~absorbers for $N_\mathrm{HI} > 10^{13.5}$ cm$^{-2}$.
        The updated TNG kinetic mode used for low black hole accretion is more gentle in terms of heating and alters the IGM thermal state to a lesser degree.
        However, TNG still does not match the COS CDD well for $N_\mathrm{HI} <10^{14}$ cm$^{-2}$.  In particular, the match of TNG to the COS data is particularly poor due to the \textit{steep slope} of the CDD. Increasing the UVB of TNG by about a factor of 2 produces a better
        fit for column densities $\rm{log N}_{HI} < 14$ because the reduced
        equilibrium ionization fraction shifts the CDDF horizontally.
        The mismatch between the predicted and observed slopes in the N$_{\rm HI} = 10^{13}$-$10^{15}$cm$^{-2}$ range in  Figure \ref{fig:mainCDD} is seemingly worse than in many other simulations \citep{Kollmeier:2014, Schaye2015MNRAS.446..521S, bolton2021limits}. We will investigate the origin of the slope mismatch in a future work.
        
        Our study  demonstrates again the key role that COS HST UV spectroscopy plays in our understanding of the low-z IGM.
        Given the time-sensitive state of HST's remaining mission, a focus on UV absorption now is critical to provide further constraints on the physical state of IGM gas covering the \Lya~transition at $0.4 < z < 1.7$, where the \Lya~forest statistics transition from dark matter/UVB dominated to baryonic feedback/UVB dominated \citep[See also,][]{Kim10.1093/mnras/staa3844}.

    \subsection{Missing turbulence in the IGM} \label{sec:bturb}
      \citet{bolton2021limits} recently suggested  that in order for the simulated and observed diffuse IGM \Lya\ $b$ distribution to match, unresolved turbulent broadening at the ratio of  $b_{\rm{turb}}$/$b_{\rm{therm}} = 0.73$ is needed.
        They found that this translates to an upper limit of $v_{\rm{turb}}=8.5$ km s$^{-1}$ ($N_{\rm{HI}}/10^{13.5}$ cm$^{-2}$).
        Observationally, non-thermal broadening can be measured via the Doppler parameters of species with different masses in the same gas phase, e.g. using OVI, CIV and HI absorbers \citep{Tripp:2008,werk2016ApJ...833...54W,burkhart2021PASP..133j2001B}.
        
        Like \citet{bolton2021limits}, we find that both Illustris and TNG produce smaller Doppler $b$ values than the COS observations.
        One possible reason for the difference in $b$ values between the simulations and observations is that the simulations are underresolving the amount of turbulence present in the IGM.
       
        However, simulations have generally found that the Doppler parameter distribution is insensitive to resolution \citep[see our analysis in the Appendix and also][]{Shull:2015}. It may be that  even higher resolution than TNG50 is required to resolve turbulent motions in the IGM and that there exists a critical resolution threshold that has not yet been reached in order to begin to see true convergence \citep{2015ApJ...805..118B}.  In addition to the issue of resolution, there may be unmodeled sources of turbulence in the IGM missing from the simulations (e.g.,  cosmic ray induced instabilities).
          In order to address these questions, including a sub-grid turbulence model for the IGM gas, similar to what has been done in disk galaxies \citep[e.g.][]{2016ApJ...826..200S} and including cosmic rays will be topics of a future exploration.

\section{Conclusions} \label{sec:con}

    In this paper we have studied the low redshift \Lya~forest statistics in the Illustris and TNG cosmological simulations in comparison to the Cosmic Origins Spectrograph (COS) UV data. 
    The TNG simulation has an identical Ultraviolet background (UVB) prescription and very similar cosmological parameters as Illustris but an entirely reworked AGN feedback prescription.  We find that:

    \begin{itemize}
  
        \item  Due to the AGN radio mode model, the original Illustris simulations have a factor of 2-3 fewer absorbers than TNG at column densities  $N_{\rm HI}< 10^{15.5}$ cm$^{-2}$. Neither TNG nor Illustris can match the slope of the COS CDD.
       
        \item The $b$ distributions in Illustris and TNG are similar, suggesting it is a poor barometer for evaluating the effects of sub-grid AGN feedback models.
        \item In agreement with \citet{bolton2021limits}, we find that both Illustris and TNG produce smaller $b$ values than the COS observations, suggesting the need for addition sources of non-thermal broadening in the low redshift IGM. 
  
        \item The amplitude and shape differences in the flux power spectrum and flux PDF between Illustris and TNG are significant (Figures \ref{fig:powerspectrum} and \ref{fig:fluxpdf}) and can not be due solely to changes in the UVB model.
        We find that the flux statistics are sensitive diagnostic constraints for sub-grid AGN feedback models.
        \item The new kinetic mode AGN feedback model in TNG substantially improves agreement with the observed flux power spectrum however TNG still overpredicts the high-k power observed in COS.

    \end{itemize}

Ultimately, resolving the discrepancies between the cosmological simulations and the HST COS data will lead  to the development of an improved UVB and AGN feedback model that is based on \textit{both the properties of galaxies \citep[e.g., ][]{Weinberger:2017} and the properties of the \Lya~forest}. 
    Developing an updated AGN feedback model based on a match of simulated and observed IGM properties and deepening our understanding of how AGN feedback alters the thermal and turbulent properties of neutral gas in the IGM will require a large campaign of numerical work that spans a range of parameters and codes. 
    Fortunately, such a simulation campaign has already been carried out as part of the CAMELS project: Cosmology and Astrophysics with MachinE Learning Simulations \citep{camels:2020}.
    CAMELS includes a suite of 2,184 state-of-the-art magnetohydrodynamic cosmological simulations performed in a volume of 25 Mpc. 
    The simulations are run either with the AREPO or GIZMO codes and employ the same baryonic subgrid physics as the TNG and SIMBA simulations \citep{2019MNRAS.486.2827D}.
    CAMELS contains thousands of different cosmological and astrophysical models by way of varying cosmological parameters and, most importantly for \Lya~forest studies at z=0.1, the parameters controlling stellar and AGN feedback and the UVB.  
    We will use these large AGN feedback parameter suites in future work constraining the nature of the effects of AGN feedback on the IGM (Tillman et al. in prep).

\appendix

\section{Resolution Effects on the CDD}

    We investigate how resolution and simulation volume size affects the \Lya~forest statistics presented above.
    
    \begin{figure*}
    \centering
    \includegraphics[width=\textwidth]{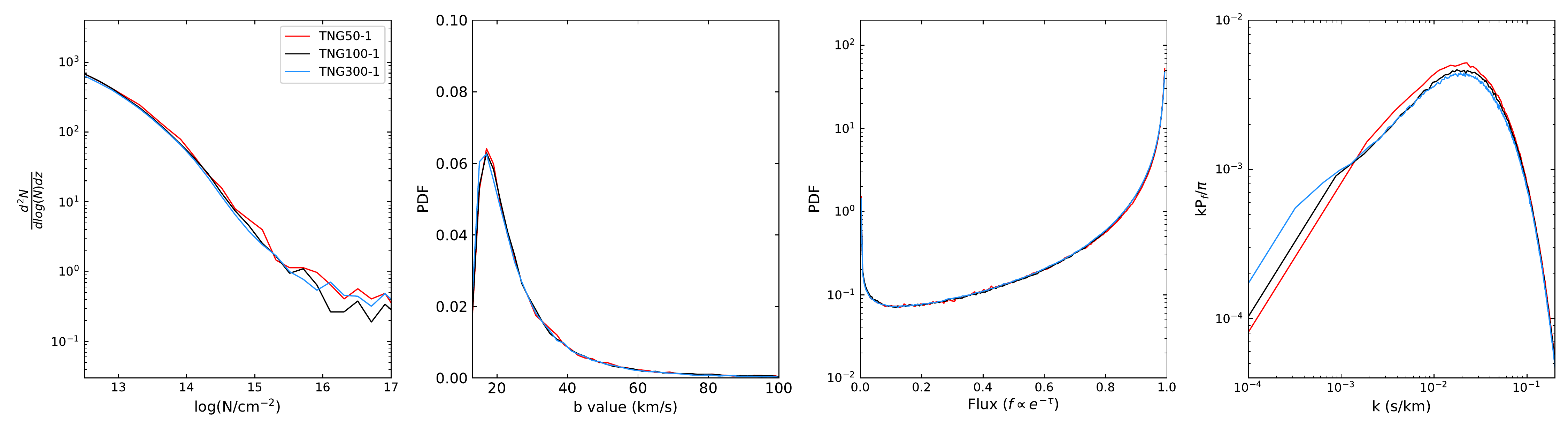}
    \caption{The various \Lya~ statistics presented in this paper for the TNG50-1, TNG100-1 and TNG300-1 simulations.  The highest resolution realizations, TNG50-1, TNG100-1 and TNG300-1, include 2$\times$2160$^3$, 2$\times$1820$^3$ and 2$\times$2500$^3$ resolution elements, respectively. The simulation box sizes go as 35, 75, and 205 Mpc h$^{-1}$ for TNG50-1, TNG100-1, and TNG300-1 respectively.}
    \label{fig:box} 
\end{figure*}
    
    Figure \ref{fig:box} presents the \Lya~ statistics of TNG50-1 (red lines), TNG100-1 (black lines) and TNG300-1 (blue lines). Each of these simulations is the highest resolution realization but they have different simulation box sizes, with TNG50-1 being the smallest volume run and TNG300-1 being the largest. 
    The left-most plot of Figure \ref{fig:box} presents the CDD. Below column densities of $10^{15}$ cm$^{-2}$ we do not find a substantial difference between the different volume runs of TNG. 
    However, higher column densities do show differences in the CDD. As expected, these high column densities represent rare over-densities around halos and TNG50 departs most significantly from TNG100 and TNG300. Our study primarily focuses on the IGM regime of column densities (i.e, below $10^{16} \rm{cm}^{-2}$). 
    Differences between the CDDs in the column density range of interest are negligible and due mostly to sample variance, demonstrating that our results are insensitive to changes in both the box size and in resolution down to 2$\times$10$^3$ resolution elements. 
    
    The second and third panels of Figure \ref{fig:box} present the Doppler parameter distributions and flux PDFs respectively. Neither of these statistics shows significant variation, implying simulation volume is not important here. The right-most panel presents the flux power spectrum. On all scales sensitivity of the flux power spectrum to resolution and volume is substantially less than both observational errors and the effect of the subgrid feedback model. Large scales ($k \lesssim 10^{-3}$ s/km) do however exhibit effects from the finite volume of the TNG50 simulation, although they are reasonably well converged for the TNG100 simulation we use for our main results. 
    
    \begin{figure*}
    \centering
    \includegraphics[width=\textwidth]{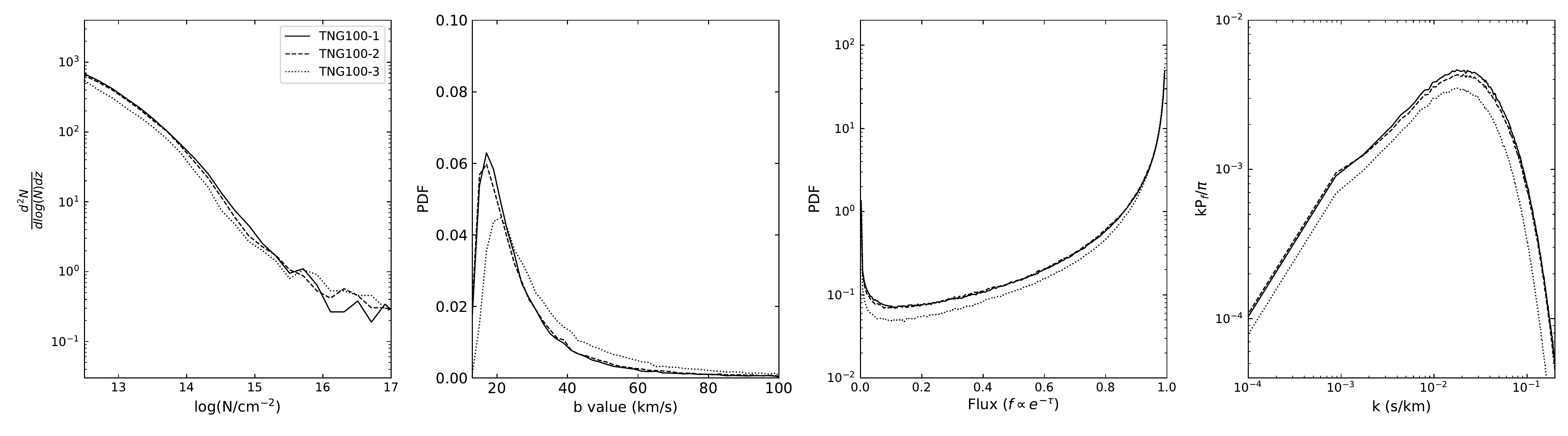}
    \caption{The \Lya statistics for different resolution elements in TNG100. 
    TNG100-1 (black solid line) has 2$\times$ 1820$^3$ resolution elements.
    TNG100-2 (black dashed line) decreases the resolution elements to  2$\times$910$^3$.
    TNG100-3 (dotted black line)  decreases the resolution elements to 2$\times$455$^3$.}
    \label{fig:res} 
\end{figure*}

\begin{figure*}
    \centering
    \includegraphics[width=\textwidth]{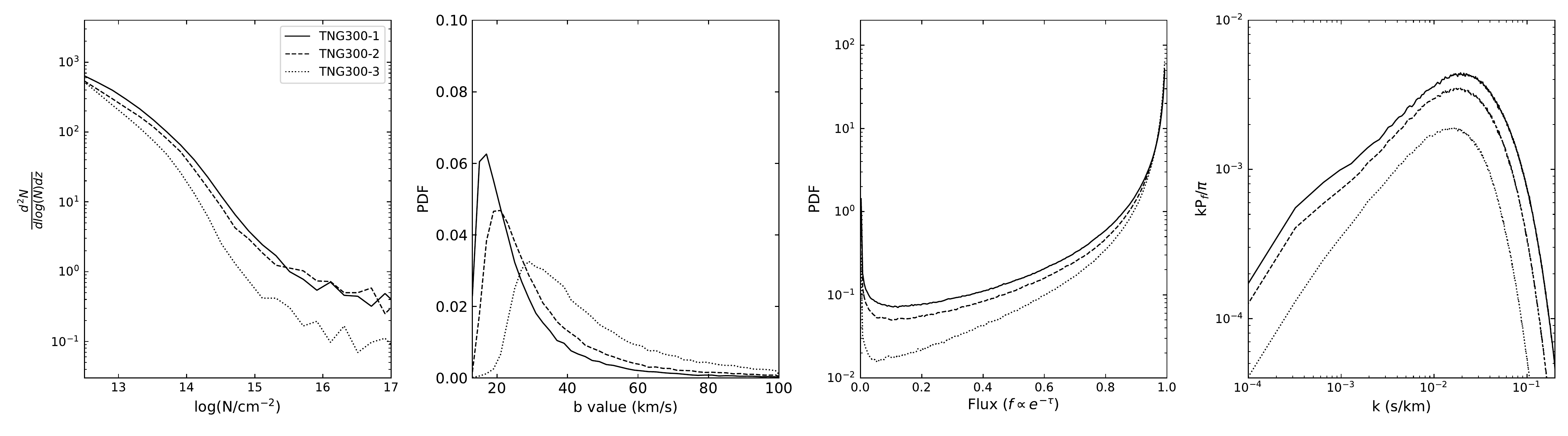}
    \caption{The \Lya~ statistics for different resolution elements in TNG300. 
    TNG300-1 (black solid line) has 2$\times$ 2500$^3$ resolution elements.
    TNG300-2 (black dashed line) decreases the resolution elements to  2$\times$1250$^3$.
    TNG300-3 (dotted black line)  decreases the resolution elements to 2$\times$625$^3$. The box size of TNG300 is  205 Mpc h$^{-1}$.}
    \label{fig:res-TNG300} 
\end{figure*}

    Figure \ref{fig:res} shows the effects of changing the number of resolution elements in TNG100.
    TNG100-1 (black solid lines) matches the original Illustris box size of 75 Mpc h$^{-1}$ and 2x 1820$^3$ resolution elements.
    TNG100-2 (black dashed lines) has the same volume size but decreases the resolution elements to  2x910$^3$.
    TNG100-3 (dotted black lines) has the same volume as TNG100-1 and TNG100-2 but decreases the resolution elements to 2x455$^3$.
    
    The left-most plot of Figure \ref{fig:res} shows the CDD for varying resolutions.
    Again, there are variations in the CDD between different resolution runs at column densities greater than log 15 cm$^{-2}$ as expected for these rare absorbers.
    TNG100-2 and TNG100-1 are fairly close, demonstrating that there is convergence in this statistic with the explored resolutions.
    TNG100-3 produces noticeable differences in the CDD across the full range of column densities, implying that the lowest resolution is not sufficient for analyzing the CDD.
    TNG100-3 produces fewer absorbers at all column densities, similar to the results for the Illustris simulation suite found in \paperone/. Convergence is similarly observed for the b-value distribution, the flux PDF, and the flux power spectrum. 
    
    While all the TNG50 and TNG100 low redshift \Lya~statistics appear to converge, the TNG300 runs statistics do not converge. Figure \ref{fig:res-TNG300} is the same as Figure \ref{fig:res} but for TNG300 runs instead. It is clear that the resolutions explored in the TNG300 runs are not sufficient to resolve the low redshift \Lya~forest statistics. The Doppler parameter distribution seems particularly sensitive to resolution.

\bibliography{ms}

\begin{thebibliography}{96}
\expandafter\ifx\csname natexlab\endcsname\relax\def\natexlab#1{#1}\fi

\bibitem[{{Altay} {et~al.}(2011){Altay}, {Theuns}, {Schaye}, {Crighton}, \&
  {Dalla Vecchia}}]{Altay2011ApJ...737L..37A}
{Altay}, G., {Theuns}, T., {Schaye}, J., {Crighton}, N. H.~M., \& {Dalla
  Vecchia}, C. 2011, \apjl, 737, L37

\bibitem[{{Bahcall} {et~al.}(1996){Bahcall}, {Bergeron}, {Boksenberg},
  {Hartig}, {Jannuzi}, {Kirhakos}, {Sargent}, {Savage}, {Schneider},
  {Turnshek}, {Weymann}, \& {Wolfe}}]{Bahcall:1996}
{Bahcall}, J.~N., {Bergeron}, J., {Boksenberg}, A., {Hartig}, G.~F., {Jannuzi},
  B.~T., {Kirhakos}, S., {Sargent}, W.~L.~W., {Savage}, B.~D., {Schneider},
  D.~P., {Turnshek}, D.~A., {Weymann}, R.~J., \& {Wolfe}, A.~M. 1996, \apj,
  457, 19

\bibitem[{{Becker} {et~al.}(2011){Becker}, {Bolton}, {Haehnelt}, \&
  {Sargent}}]{Becker:2011}
{Becker}, G.~D., {Bolton}, J.~S., {Haehnelt}, M.~G., \& {Sargent}, W. L.~W.
  2011, \mnras, 410, 1096

\bibitem[{{Bird}(2017)}]{Bird:2017}
{Bird}, S. 2017, {FSFE: Fake Spectra Flux Extractor}

\bibitem[{{Bird} {et~al.}(2015){Bird}, {Haehnelt}, {Neeleman}, {Genel},
  {Vogelsberger}, \& {Hernquist}}]{Bird:2015}
{Bird}, S., {Haehnelt}, M., {Neeleman}, M., {Genel}, S., {Vogelsberger}, M., \&
  {Hernquist}, L. 2015, MNRAS, 447, 1834

\bibitem[{{Bird} {et~al.}(2014){Bird}, {Vogelsberger}, {Haehnelt}, {Sijacki},
  {Genel}, {Torrey}, {Springel}, \& {Hernquist}}]{Bird:2014}
{Bird}, S., {Vogelsberger}, M., {Haehnelt}, M., {Sijacki}, D., {Genel}, S.,
  {Torrey}, P., {Springel}, V., \& {Hernquist}, L. 2014, \mnras, 445, 2313

\bibitem[{{Boera} {et~al.}(2014){Boera}, {Murphy}, {Becker}, \&
  {Bolton}}]{Boera:2014}
{Boera}, E., {Murphy}, M.~T., {Becker}, G.~D., \& {Bolton}, J.~S. 2014, MNRAS,
  441, 1916

\bibitem[{Bolton {et~al.}(2021)Bolton, Gaikwad, Haehnelt, Kim, Nasir, Puchwein,
  Viel, \& Wakker}]{bolton2021limits}
Bolton, J.~S., Gaikwad, P., Haehnelt, M.~G., Kim, T.-S., Nasir, F., Puchwein,
  E., Viel, M., \& Wakker, B.~P. 2021, Limits on non-canonical heating and
  turbulence in the intergalactic medium from the low redshift Lyman-alpha
  forest

\bibitem[{{Borrow} {et~al.}(2020){Borrow}, {Angl{\'e}s-Alc{\'a}zar}, \&
  {Dav{\'e}}}]{borrow2020MNRAS.491.6102B}
{Borrow}, J., {Angl{\'e}s-Alc{\'a}zar}, D., \& {Dav{\'e}}, R. 2020, \mnras,
  491, 6102

\bibitem[{{Burkhart}(2021)}]{burkhart2021PASP..133j2001B}
{Burkhart}, B. 2021, \pasp, 133, 102001

\bibitem[{{Burkhart} {et~al.}(2015){Burkhart}, {Lazarian}, {Balsara}, {Meyer},
  \& {Cho}}]{2015ApJ...805..118B}
{Burkhart}, B., {Lazarian}, A., {Balsara}, D., {Meyer}, C., \& {Cho}, J. 2015,
  \apj, 805, 118

\bibitem[{{Cen} {et~al.}(1994){Cen}, {Miralda-Escud{\'e}}, {Ostriker}, \&
  {Rauch}}]{Cen:1994}
{Cen}, R., {Miralda-Escud{\'e}}, J., {Ostriker}, J.~P., \& {Rauch}, M. 1994,
  ApJ, 437, L9

\bibitem[{{Chabanier} {et~al.}(2020){Chabanier}, {Bournaud}, {Dubois},
  {Palanque-Delabrouille}, {Y{\`e}che}, {Armengaud}, {Peirani}, \&
  {Beckmann}}]{2020MNRAS.495.1825C}
{Chabanier}, S., {Bournaud}, F., {Dubois}, Y., {Palanque-Delabrouille}, N.,
  {Y{\`e}che}, C., {Armengaud}, E., {Peirani}, S., \& {Beckmann}, R. 2020,
  \mnras, 495, 1825

\bibitem[{{Chang} {et~al.}(2012){Chang}, {Broderick}, \&
  {Pfrommer}}]{Chang:2012}
{Chang}, P., {Broderick}, A.~E., \& {Pfrommer}, C. 2012, \apj, 752, 23

\bibitem[{{Christiansen} {et~al.}(2020){Christiansen}, {Dav{\'e}}, {Sorini}, \&
  {Angl{\'e}s-Alc{\'a}zar}}]{Christiansen2020MNRAS.499.2617C}
{Christiansen}, J.~F., {Dav{\'e}}, R., {Sorini}, D., \&
  {Angl{\'e}s-Alc{\'a}zar}, D. 2020, \mnras, 499, 2617

\bibitem[{{Croft} {et~al.}(1998){Croft}, {Weinberg}, {Katz}, \&
  {Hernquist}}]{Croft:1998}
{Croft}, R. A.~C., {Weinberg}, D.~H., {Katz}, N., \& {Hernquist}, L. 1998,
  \apj, 495, 44

\bibitem[{{Danforth} {et~al.}(2016){Danforth}, {Keeney}, {Tilton}, {Shull},
  {Stocke}, {Stevans}, {Pieri}, {Savage}, {France}, {Syphers}, {Smith},
  {Green}, {Froning}, {Penton}, \& {Osterman}}]{Danforth:2016}
{Danforth}, C.~W., {Keeney}, B.~A., {Tilton}, E.~M., {Shull}, J.~M., {Stocke},
  J.~T., {Stevans}, M., {Pieri}, M.~M., {Savage}, B.~D., {France}, K.,
  {Syphers}, D., {Smith}, B.~D., {Green}, J.~C., {Froning}, C., {Penton},
  S.~V., \& {Osterman}, S.~N. 2016, ApJ, 817, 111

\bibitem[{{Danforth} \& {Shull}(2005)}]{Danforth:2005}
{Danforth}, C.~W. \& {Shull}, J.~M. 2005, \apj, 624, 555

\bibitem[{{Danforth} {et~al.}(2010){Danforth}, {Stocke}, \&
  {Shull}}]{Danforth:2010}
{Danforth}, C.~W., {Stocke}, J.~T., \& {Shull}, J.~M. 2010, \apj, 710, 613

\bibitem[{{Dav{\'e}} {et~al.}(2019){Dav{\'e}}, {Angl{\'e}s-Alc{\'a}zar},
  {Narayanan}, {Li}, {Rafieferantsoa}, \& {Appleby}}]{2019MNRAS.486.2827D}
{Dav{\'e}}, R., {Angl{\'e}s-Alc{\'a}zar}, D., {Narayanan}, D., {Li}, Q.,
  {Rafieferantsoa}, M.~H., \& {Appleby}, S. 2019, \mnras, 486, 2827

\bibitem[{{Dav{\'e}} {et~al.}(2001){Dav{\'e}}, {Cen}, {Ostriker}, {Bryan},
  {Hernquist}, {Katz}, {Weinberg}, {Norman}, \& {O'Shea}}]{Dave:2001}
{Dav{\'e}}, R., {Cen}, R., {Ostriker}, J.~P., {Bryan}, G.~L., {Hernquist}, L.,
  {Katz}, N., {Weinberg}, D.~H., {Norman}, M.~L., \& {O'Shea}, B. 2001, \apj,
  552, 473

\bibitem[{{Dav{\'e}} {et~al.}(1999){Dav{\'e}}, {Hernquist}, {Katz}, \&
  {Weinberg}}]{Dave:1999}
{Dav{\'e}}, R., {Hernquist}, L., {Katz}, N., \& {Weinberg}, D.~H. 1999, \apj,
  511, 521

\bibitem[{{Faucher-Gigu{\`e}re}(2020)}]{Faucher2020MNRAS.493.1614F}
{Faucher-Gigu{\`e}re}, C.-A. 2020, \mnras, 493, 1614

\bibitem[{{Faucher-Gigu{\`e}re} {et~al.}(2008){Faucher-Gigu{\`e}re}, {Lidz},
  {Hernquist}, \& {Zaldarriaga}}]{Faucher-Giguere2008}
{Faucher-Gigu{\`e}re}, C.-A., {Lidz}, A., {Hernquist}, L., \& {Zaldarriaga}, M.
  2008, \apj, 688, 85

\bibitem[{{Faucher-Gigu{\`e}re} {et~al.}(2009){Faucher-Gigu{\`e}re}, {Lidz},
  {Zaldarriaga}, \& {Hernquist}}]{Faucher-Giguere:2009}
{Faucher-Gigu{\`e}re}, C.-A., {Lidz}, A., {Zaldarriaga}, M., \& {Hernquist}, L.
  2009, ApJ, 703, 1416

\bibitem[{{Gaikwad} {et~al.}(2017){Gaikwad}, {Khaire}, {Choudhury}, \&
  {Srianand}}]{Gaikwad:2017}
{Gaikwad}, P., {Khaire}, V., {Choudhury}, T.~R., \& {Srianand}, R. 2017, MNRAS,
  466, 838

\bibitem[{{Genel} {et~al.}(2014){Genel}, {Vogelsberger}, {Springel}, {Sijacki},
  {Nelson}, {Snyder}, {Rodriguez-Gomez}, {Torrey}, \& {Hernquist}}]{Genel:2014}
{Genel}, S., {Vogelsberger}, M., {Springel}, V., {Sijacki}, D., {Nelson}, D.,
  {Snyder}, G., {Rodriguez-Gomez}, V., {Torrey}, P., \& {Hernquist}, L. 2014,
  \mnras, 445, 175

\bibitem[{{Gunn} \& {Peterson}(1965)}]{Gunn:1965}
{Gunn}, J.~E. \& {Peterson}, B.~A. 1965, ApJ, 142, 1633

\bibitem[{{Gurvich} {et~al.}(2017){Gurvich}, {Burkhart}, \&
  {Bird}}]{Gurvich2017}
{Gurvich}, A., {Burkhart}, B., \& {Bird}, S. 2017, ApJ, 835, 175

\bibitem[{{Haardt} \& {Madau}(1996)}]{Haardt:1996}
{Haardt}, F. \& {Madau}, P. 1996, \apj, 461, 20

\bibitem[{{Haardt} \& {Madau}(2012)}]{Haardt:2012}
---. 2012, \apj, 746, 125

\bibitem[{{Henden} {et~al.}(2018){Henden}, {Puchwein}, {Shen}, \&
  {Sijacki}}]{henden2018MNRAS.479.5385H}
{Henden}, N.~A., {Puchwein}, E., {Shen}, S., \& {Sijacki}, D. 2018, \mnras,
  479, 5385

\bibitem[{{Hernquist} {et~al.}(1996){Hernquist}, {Katz}, {Weinberg}, \&
  {Miralda-Escud{\'e}}}]{Hernquist:1996}
{Hernquist}, L., {Katz}, N., {Weinberg}, D.~H., \& {Miralda-Escud{\'e}}, J.
  1996, \apj, 457, L51

\bibitem[{{Hinshaw} {et~al.}(2013){Hinshaw}, {Larson}, {Komatsu}, {Spergel},
  {Bennett}, {Dunkley}, {Nolta}, {Halpern}, {Hill}, {Odegard}, {Page}, {Smith},
  {Weiland}, {Gold}, {Jarosik}, {Kogut}, {Limon}, {Meyer}, {Tucker}, {Wollack},
  \& {Wright}}]{2013ApJS..208...19H}
{Hinshaw}, G., {Larson}, D., {Komatsu}, E., {Spergel}, D.~N., {Bennett}, C.~L.,
  {Dunkley}, J., {Nolta}, M.~R., {Halpern}, M., {Hill}, R.~S., {Odegard}, N.,
  {Page}, L., {Smith}, K.~M., {Weiland}, J.~L., {Gold}, B., {Jarosik}, N.,
  {Kogut}, A., {Limon}, M., {Meyer}, S.~S., {Tucker}, G.~S., {Wollack}, E., \&
  {Wright}, E.~L. 2013, \apjs, 208, 19

\bibitem[{{Hiss} {et~al.}(2018){Hiss}, {Walther}, {Hennawi}, {O{\~n}orbe},
  {O'Meara}, {Rorai}, \& {Luki{\'c}}}]{Hiss2018ApJ...865...42H}
{Hiss}, H., {Walther}, M., {Hennawi}, J.~F., {O{\~n}orbe}, J., {O'Meara},
  J.~M., {Rorai}, A., \& {Luki{\'c}}, Z. 2018, \apj, 865, 42

\bibitem[{{Hui} \& {Gnedin}(1997)}]{1997MNRAS.292...27H}
{Hui}, L. \& {Gnedin}, N.~Y. 1997, \mnras, 292, 27

\bibitem[{{Jannuzi} {et~al.}(1998){Jannuzi}, {Bahcall}, {Bergeron},
  {Boksenberg}, {Hartig}, {Kirhakos}, {Sargent}, {Savage}, {Schneider},
  {Turnshek}, {Weymann}, \& {Wolfe}}]{Jannuzi:1998}
{Jannuzi}, B.~T., {Bahcall}, J.~N., {Bergeron}, J., {Boksenberg}, A., {Hartig},
  G.~F., {Kirhakos}, S., {Sargent}, W.~L.~W., {Savage}, B.~D., {Schneider},
  D.~P., {Turnshek}, D.~A., {Weymann}, R.~J., \& {Wolfe}, A.~M. 1998, \apjs,
  118, 1

\bibitem[{{Katz} {et~al.}(1996){Katz}, {Weinberg}, \& {Hernquist}}]{Katz:1996}
{Katz}, N., {Weinberg}, D.~H., \& {Hernquist}, L. 1996, The Astrophysical
  Journal Supplement Series, 105, 19

\bibitem[{{Khaire} \& {Srianand}(2015)}]{Khaire:2015}
{Khaire}, V. \& {Srianand}, R. 2015, \mnras, 451, L30

\bibitem[{{Khaire} \& {Srianand}(2019)}]{khaireSri2019MNRAS.484.4174K}
---. 2019, \mnras, 484, 4174

\bibitem[{{Khaire} {et~al.}(2019){Khaire}, {Walther}, {Hennawi}, {O{\~n}orbe},
  {Luki{\'c}}, {}, {Prochaska}, {Tripp}, {Burchett}, \&
  {Rodriguez}}]{Khaire2019}
{Khaire}, V., {Walther}, M., {Hennawi}, J.~F., {O{\~n}orbe}, J., {Luki{\'c}},
  {}, Z., {Prochaska}, J.~X., {Tripp}, T.~M., {Burchett}, J.~N., \&
  {Rodriguez}, C. 2019, \mnras, 486, 769

\bibitem[{Kim {et~al.}(2020)Kim, Wakker, Nasir, Carswell, Savage, Bolton, Fox,
  Viel, Haehnelt, Charlton, \& Rosenwasser}]{Kim10.1093/mnras/staa3844}
Kim, T.-S., Wakker, B.~P., Nasir, F., Carswell, R.~F., Savage, B.~D., Bolton,
  J.~S., Fox, A.~J., Viel, M., Haehnelt, M.~G., Charlton, J.~C., \&
  Rosenwasser, B.~E. 2020, Monthly Notices of the Royal Astronomical Society,
  501, 5811

\bibitem[{{Kollmeier} {et~al.}(2014){Kollmeier}, {Weinberg}, {Oppenheimer},
  {Haardt}, {Katz}, {Dav{\'e}}, {Fardal}, {Madau}, {Danforth}, {Ford},
  {Peeples}, \& {McEwen}}]{Kollmeier:2014}
{Kollmeier}, J.~A., {Weinberg}, D.~H., {Oppenheimer}, B.~D., {Haardt}, F.,
  {Katz}, N., {Dav{\'e}}, R., {Fardal}, M., {Madau}, P., {Danforth}, C.,
  {Ford}, A.~B., {Peeples}, M.~S., \& {McEwen}, J. 2014, ApJ, 789, L32

\bibitem[{{Lehner} {et~al.}(2007){Lehner}, {Savage}, {Richter}, {Sembach},
  {Tripp}, \& {Wakker}}]{Lehner:2007}
{Lehner}, N., {Savage}, B.~D., {Richter}, P., {Sembach}, K.~R., {Tripp}, T.~M.,
  \& {Wakker}, B.~P. 2007, \apj, 658, 680

\bibitem[{{Lidz} {et~al.}(2010){Lidz}, {Faucher-Gigu{\`e}re}, {Dall'Aglio},
  {McQuinn}, {Fechner}, {Zaldarriaga}, {Hernquist}, \& {Dutta}}]{Lidz:2010}
{Lidz}, A., {Faucher-Gigu{\`e}re}, C.-A., {Dall'Aglio}, A., {McQuinn}, M.,
  {Fechner}, C., {Zaldarriaga}, M., {Hernquist}, L., \& {Dutta}, S. 2010, \apj,
  718, 199

\bibitem[{{Marinacci} {et~al.}(2018){Marinacci}, {Vogelsberger}, {Kannan},
  {Mocz}, {Pakmor}, \& {Springel}}]{Marinacci:2018}
{Marinacci}, F., {Vogelsberger}, M., {Kannan}, R., {Mocz}, P., {Pakmor}, R., \&
  {Springel}, V. 2018, \mnras, 476, 2476

\bibitem[{{Martizzi} {et~al.}(2019){Martizzi}, {Vogelsberger}, {Artale},
  {Haider}, {Torrey}, {Marinacci}, {Nelson}, {Pillepich}, {Weinberger},
  {Hernquist}, {Naiman}, \& {Springel}}]{Martizzi2019MNRAS.486.3766M}
{Martizzi}, D., {Vogelsberger}, M., {Artale}, M.~C., {Haider}, M., {Torrey},
  P., {Marinacci}, F., {Nelson}, D., {Pillepich}, A., {Weinberger}, R.,
  {Hernquist}, L., {Naiman}, J., \& {Springel}, V. 2019, \mnras, 486, 3766

\bibitem[{{McDonald} {et~al.}(2005){McDonald}, {Seljak}, {Cen}, {Shih},
  {Weinberg}, {Burles}, {Schneider}, {Schlegel}, {Bahcall}, {Briggs},
  {Brinkmann}, {Fukugita}, {Ivezi{\'c}}, {Kent}, \& {Vanden
  Berk}}]{McDonald:2005}
{McDonald}, P., {Seljak}, U., {Cen}, R., {Shih}, D., {Weinberg}, D.~H.,
  {Burles}, S., {Schneider}, D.~P., {Schlegel}, D.~J., {Bahcall}, N.~A.,
  {Briggs}, J.~W., {Brinkmann}, J., {Fukugita}, M., {Ivezi{\'c}}, {\v{Z}}.,
  {Kent}, S., \& {Vanden Berk}, D.~E. 2005, \apj, 635, 761

\bibitem[{{McQuinn}(2016)}]{McQuinn2016ARA&A..54..313M}
{McQuinn}, M. 2016, \araa, 54, 313

\bibitem[{{Meiring} {et~al.}(2011){Meiring}, {Tripp}, {Prochaska}, {Tumlinson},
  {Werk}, {Jenkins}, {Thom}, {O'Meara}, \&
  {Sembach}}]{Meiring2011ApJ...732...35M}
{Meiring}, J.~D., {Tripp}, T.~M., {Prochaska}, J.~X., {Tumlinson}, J., {Werk},
  J., {Jenkins}, E.~B., {Thom}, C., {O'Meara}, J.~M., \& {Sembach}, K.~R. 2011,
  \apj, 732, 35

\bibitem[{{Miralda-Escud{\'e}} {et~al.}(1996){Miralda-Escud{\'e}}, {Cen},
  {Ostriker}, \& {Rauch}}]{Miralda-Escude:1996}
{Miralda-Escud{\'e}}, J., {Cen}, R., {Ostriker}, J.~P., \& {Rauch}, M. 1996,
  \apj, 471, 582

\bibitem[{{Naiman} {et~al.}(2018){Naiman}, {Pillepich}, {Springel},
  {Ramirez-Ruiz}, {Torrey}, {Vogelsberger}, {Pakmor}, {Nelson}, {Marinacci},
  {Hernquist}, {Weinberger}, \& {Genel}}]{Naiman2018MNRAS.477.1206N}
{Naiman}, J.~P., {Pillepich}, A., {Springel}, V., {Ramirez-Ruiz}, E., {Torrey},
  P., {Vogelsberger}, M., {Pakmor}, R., {Nelson}, D., {Marinacci}, F.,
  {Hernquist}, L., {Weinberger}, R., \& {Genel}, S. 2018, \mnras, 477, 1206

\bibitem[{{Nasir} {et~al.}(2017){Nasir}, {Bolton}, {Viel}, {Kim}, {Haehnelt},
  {Puchwein}, \& {Sijacki}}]{Nasir:2017}
{Nasir}, F., {Bolton}, J.~S., {Viel}, M., {Kim}, T.-S., {Haehnelt}, M.~G.,
  {Puchwein}, E., \& {Sijacki}, D. 2017, MNRAS, 471, 1056

\bibitem[{{Nelson} {et~al.}(2015){Nelson}, {Pillepich}, {Genel},
  {Vogelsberger}, {Springel}, {Torrey}, {Rodriguez-Gomez}, {Sijacki}, {Snyder},
  {Griffen}, {Marinacci}, {Blecha}, {Sales}, {Xu}, \&
  {Hernquist}}]{Nelson:2015}
{Nelson}, D., {Pillepich}, A., {Genel}, S., {Vogelsberger}, M., {Springel}, V.,
  {Torrey}, P., {Rodriguez-Gomez}, V., {Sijacki}, D., {Snyder}, G.~F.,
  {Griffen}, B., {Marinacci}, F., {Blecha}, L., {Sales}, L., {Xu}, D., \&
  {Hernquist}, L. 2015, Astronomy and Computing, 13, 12

\bibitem[{{Nelson} {et~al.}(2019{\natexlab{a}}){Nelson}, {Pillepich},
  {Springel}, {Pakmor}, {Weinberger}, {Genel}, {Torrey}, {Vogelsberger},
  {Marinacci}, \& {Hernquist}}]{Nelson:2019a}
{Nelson}, D., {Pillepich}, A., {Springel}, V., {Pakmor}, R., {Weinberger}, R.,
  {Genel}, S., {Torrey}, P., {Vogelsberger}, M., {Marinacci}, F., \&
  {Hernquist}, L. 2019{\natexlab{a}}, \mnras, 490, 3234

\bibitem[{{Nelson} {et~al.}(2018){Nelson}, {Pillepich}, {Springel},
  {Weinberger}, {Hernquist}, {Pakmor}, {Genel}, {Torrey}, {Vogelsberger}, \&
  {Kauffmann}}]{Nelson2018}
{Nelson}, D., {Pillepich}, A., {Springel}, V., {Weinberger}, R., {Hernquist},
  L., {Pakmor}, R., {Genel}, S., {Torrey}, P., {Vogelsberger}, M., \&
  {Kauffmann}, G. 2018, \mnras, 475, 624

\bibitem[{{Nelson} {et~al.}(2019{\natexlab{b}}){Nelson}, {Springel},
  {Pillepich}, {Rodriguez-Gomez}, {Torrey}, {Genel}, {Vogelsberger}, {Pakmor},
  {Marinacci}, {Weinberger}, {Kelley}, {Lovell}, {Diemer}, \&
  {Hernquist}}]{Nelson:2019b}
{Nelson}, D., {Springel}, V., {Pillepich}, A., {Rodriguez-Gomez}, V., {Torrey},
  P., {Genel}, S., {Vogelsberger}, M., {Pakmor}, R., {Marinacci}, F.,
  {Weinberger}, R., {Kelley}, L., {Lovell}, M., {Diemer}, B., \& {Hernquist},
  L. 2019{\natexlab{b}}, Computational Astrophysics and Cosmology, 6, 2

\bibitem[{{Palanque-Delabrouille} {et~al.}(2013){Palanque-Delabrouille},
  {Y{\`e}che}, {Borde}, {Le Goff}, {Rossi}, {Viel}, {Aubourg}, {Bailey},
  {Bautista}, {Blomqvist}, {Bolton}, {Bolton}, {Busca}, {Carithers}, {Croft},
  {Dawson}, {Delubac}, {Font-Ribera}, {Ho}, {Kirkby}, {Lee}, {Margala},
  {Miralda-Escud{\'e}}, {Muna}, {Myers}, {Noterdaeme}, {P{\^a}ris},
  {Petitjean}, {Pieri}, {Rich}, {Rollinde}, {Ross}, {Schlegel}, {Schneider},
  {Slosar}, \& {Weinberg}}]{Palanque-Delabrouille:2013}
{Palanque-Delabrouille}, N., {Y{\`e}che}, C., {Borde}, A., {Le Goff}, J.-M.,
  {Rossi}, G., {Viel}, M., {Aubourg}, {\'E}., {Bailey}, S., {Bautista}, J.,
  {Blomqvist}, M., {Bolton}, A., {Bolton}, J.~S., {Busca}, N.~G., {Carithers},
  B., {Croft}, R.~A.~C., {Dawson}, K.~S., {Delubac}, T., {Font-Ribera}, A.,
  {Ho}, S., {Kirkby}, D., {Lee}, K.-G., {Margala}, D., {Miralda-Escud{\'e}},
  J., {Muna}, D., {Myers}, A.~D., {Noterdaeme}, P., {P{\^a}ris}, I.,
  {Petitjean}, P., {Pieri}, M.~M., {Rich}, J., {Rollinde}, E., {Ross}, N.~P.,
  {Schlegel}, D.~J., {Schneider}, D.~P., {Slosar}, A., \& {Weinberg}, D.~H.
  2013, AAP, 559, A85

\bibitem[{{Penton} {et~al.}(2000){Penton}, {Stocke}, \& {Shull}}]{Penton:2000}
{Penton}, S.~V., {Stocke}, J.~T., \& {Shull}, J.~M. 2000, \apjs, 130, 121

\bibitem[{{Penton} {et~al.}(2004){Penton}, {Stocke}, \& {Shull}}]{Penton:2004}
---. 2004, \apjs, 152, 29

\bibitem[{{Pillepich} {et~al.}(2018{\natexlab{a}}){Pillepich}, {Nelson},
  {Hernquist}, {Springel}, {Pakmor}, {Torrey}, {Weinberger}, {Genel}, {Naiman},
  {Marinacci}, \& {Vogelsberger}}]{Pillepich2018MNRAS.475..648P}
{Pillepich}, A., {Nelson}, D., {Hernquist}, L., {Springel}, V., {Pakmor}, R.,
  {Torrey}, P., {Weinberger}, R., {Genel}, S., {Naiman}, J.~P., {Marinacci},
  F., \& {Vogelsberger}, M. 2018{\natexlab{a}}, \mnras, 475, 648

\bibitem[{{Pillepich} {et~al.}(2019){Pillepich}, {Nelson}, {Springel},
  {Pakmor}, {Torrey}, {Weinberger}, {Vogelsberger}, {Marinacci}, {Genel}, {van
  der Wel}, \& {Hernquist}}]{Pillepich:2019}
{Pillepich}, A., {Nelson}, D., {Springel}, V., {Pakmor}, R., {Torrey}, P.,
  {Weinberger}, R., {Vogelsberger}, M., {Marinacci}, F., {Genel}, S., {van der
  Wel}, A., \& {Hernquist}, L. 2019, \mnras, 490, 3196

\bibitem[{{Pillepich} {et~al.}(2018{\natexlab{b}}){Pillepich}, {Springel},
  {Nelson}, {Genel}, {Naiman}, {Pakmor}, {Hernquist}, {Torrey}, {Vogelsberger},
  {Weinberger}, \& {Marinacci}}]{Pillepich:2018b}
{Pillepich}, A., {Springel}, V., {Nelson}, D., {Genel}, S., {Naiman}, J.,
  {Pakmor}, R., {Hernquist}, L., {Torrey}, P., {Vogelsberger}, M.,
  {Weinberger}, R., \& {Marinacci}, F. 2018{\natexlab{b}}, \mnras, 473, 4077

\bibitem[{{Planck Collaboration} {et~al.}(2016){Planck Collaboration}, {Ade},
  {Aghanim}, {Arnaud}, {Ashdown}, {Aumont}, {Baccigalupi}, {Banday},
  {Barreiro}, {Bartlett}, {Bartolo}, {Battaner}, {Battye}, {Benabed},
  {Beno{\^\i}t}, {Benoit-L{\'e}vy}, {Bernard}, {Bersanelli}, {Bielewicz},
  {Bock}, {Bonaldi}, {Bonavera}, {Bond}, {Borrill}, {Bouchet}, {Boulanger},
  {Bucher}, {Burigana}, {Butler}, {Calabrese}, {Cardoso}, {Catalano},
  {Challinor}, {Chamballu}, {Chary}, {Chiang}, {Chluba}, {Christensen},
  {Church}, {Clements}, {Colombi}, {Colombo}, {Combet}, {Coulais}, {Crill},
  {Curto}, {Cuttaia}, {Danese}, {Davies}, {Davis}, {de Bernardis}, {de Rosa},
  {de Zotti}, {Delabrouille}, {D{\'e}sert}, {Di Valentino}, {Dickinson},
  {Diego}, {Dolag}, {Dole}, {Donzelli}, {Dor{\'e}}, {Douspis}, {Ducout},
  {Dunkley}, {Dupac}, {Efstathiou}, {Elsner}, {En{\ss}lin}, {Eriksen},
  {Farhang}, {Fergusson}, {Finelli}, {Forni}, {Frailis}, {Fraisse},
  {Franceschi}, {Frejsel}, {Galeotta}, {Galli}, {Ganga}, {Gauthier}, {Gerbino},
  {Ghosh}, {Giard}, {Giraud-H{\'e}raud}, {Giusarma}, {Gjerl{\o}w},
  {Gonz{\'a}lez-Nuevo}, {G{\'o}rski}, {Gratton}, {Gregorio}, {Gruppuso},
  {Gudmundsson}, {Hamann}, {Hansen}, {Hanson}, {Harrison}, {Helou},
  {Henrot-Versill{\'e}}, {Hern{\'a}ndez-Monteagudo}, {Herranz}, {Hildebrandt},
  {Hivon}, {Hobson}, {Holmes}, {Hornstrup}, {Hovest}, {Huang}, {Huffenberger},
  {Hurier}, {Jaffe}, {Jaffe}, {Jones}, {Juvela}, {Keih{\"a}nen}, {Keskitalo},
  {Kisner}, {Kneissl}, {Knoche}, {Knox}, {Kunz}, {Kurki-Suonio}, {Lagache},
  {L{\"a}hteenm{\"a}ki}, {Lamarre}, {Lasenby}, {Lattanzi}, {Lawrence}, {Leahy},
  {Leonardi}, {Lesgourgues}, {Levrier}, {Lewis}, {Liguori}, {Lilje},
  {Linden-V{\o}rnle}, {L{\'o}pez-Caniego}, {Lubin}, {Mac{\'\i}as-P{\'e}rez},
  {Maggio}, {Maino}, {Mandolesi}, {Mangilli}, {Marchini}, {Maris}, {Martin},
  {Martinelli}, {Mart{\'\i}nez-Gonz{\'a}lez}, {Masi}, {Matarrese}, {McGehee},
  {Meinhold}, {Melchiorri}, {Melin}, {Mendes}, {Mennella}, {Migliaccio},
  {Millea}, {Mitra}, {Miville-Desch{\^e}nes}, {Moneti}, {Montier}, {Morgante},
  {Mortlock}, {Moss}, {Munshi}, {Murphy}, {Naselsky}, {Nati}, {Natoli},
  {Netterfield}, {N{\o}rgaard-Nielsen}, {Noviello}, {Novikov}, {Novikov},
  {Oxborrow}, {Paci}, {Pagano}, {Pajot}, {Paladini}, {Paoletti}, {Partridge},
  {Pasian}, {Patanchon}, {Pearson}, {Perdereau}, {Perotto}, {Perrotta},
  {Pettorino}, {Piacentini}, {Piat}, {Pierpaoli}, {Pietrobon}, {Plaszczynski},
  {Pointecouteau}, {Polenta}, {Popa}, {Pratt}, {Pr{\'e}zeau}, {Prunet},
  {Puget}, {Rachen}, {Reach}, {Rebolo}, {Reinecke}, {Remazeilles}, {Renault},
  {Renzi}, {Ristorcelli}, {Rocha}, {Rosset}, {Rossetti}, {Roudier},
  {Rouill{\'e} d'Orfeuil}, {Rowan-Robinson}, {Rubi{\~n}o-Mart{\'\i}n},
  {Rusholme}, {Said}, {Salvatelli}, {Salvati}, {Sandri}, {Santos},
  {Savelainen}, {Savini}, {Scott}, {Seiffert}, {Serra}, {Shellard}, {Spencer},
  {Spinelli}, {Stolyarov}, {Stompor}, {Sudiwala}, {Sunyaev}, {Sutton},
  {Suur-Uski}, {Sygnet}, {Tauber}, {Terenzi}, {Toffolatti}, {Tomasi},
  {Tristram}, {Trombetti}, {Tucci}, {Tuovinen}, {T{\"u}rler}, {Umana},
  {Valenziano}, {Valiviita}, {Van Tent}, {Vielva}, {Villa}, {Wade}, {Wandelt},
  {Wehus}, {White}, {White}, {Wilkinson}, {Yvon}, {Zacchei}, \&
  {Zonca}}]{2016A&A...594A..13P}
{Planck Collaboration}, {Ade}, P.~A.~R., {Aghanim}, N., {Arnaud}, M.,
  {Ashdown}, M., {Aumont}, J., {Baccigalupi}, C., {Banday}, A.~J., {Barreiro},
  R.~B., {Bartlett}, J.~G., {Bartolo}, N., {Battaner}, E., {Battye}, R.,
  {Benabed}, K., {Beno{\^\i}t}, A., {Benoit-L{\'e}vy}, A., {Bernard}, J.~P.,
  {Bersanelli}, M., {Bielewicz}, P., {Bock}, J.~J., {Bonaldi}, A., {Bonavera},
  L., {Bond}, J.~R., {Borrill}, J., {Bouchet}, F.~R., {Boulanger}, F.,
  {Bucher}, M., {Burigana}, C., {Butler}, R.~C., {Calabrese}, E., {Cardoso},
  J.~F., {Catalano}, A., {Challinor}, A., {Chamballu}, A., {Chary}, R.~R.,
  {Chiang}, H.~C., {Chluba}, J., {Christensen}, P.~R., {Church}, S.,
  {Clements}, D.~L., {Colombi}, S., {Colombo}, L.~P.~L., {Combet}, C.,
  {Coulais}, A., {Crill}, B.~P., {Curto}, A., {Cuttaia}, F., {Danese}, L.,
  {Davies}, R.~D., {Davis}, R.~J., {de Bernardis}, P., {de Rosa}, A., {de
  Zotti}, G., {Delabrouille}, J., {D{\'e}sert}, F.~X., {Di Valentino}, E.,
  {Dickinson}, C., {Diego}, J.~M., {Dolag}, K., {Dole}, H., {Donzelli}, S.,
  {Dor{\'e}}, O., {Douspis}, M., {Ducout}, A., {Dunkley}, J., {Dupac}, X.,
  {Efstathiou}, G., {Elsner}, F., {En{\ss}lin}, T.~A., {Eriksen}, H.~K.,
  {Farhang}, M., {Fergusson}, J., {Finelli}, F., {Forni}, O., {Frailis}, M.,
  {Fraisse}, A.~A., {Franceschi}, E., {Frejsel}, A., {Galeotta}, S., {Galli},
  S., {Ganga}, K., {Gauthier}, C., {Gerbino}, M., {Ghosh}, T., {Giard}, M.,
  {Giraud-H{\'e}raud}, Y., {Giusarma}, E., {Gjerl{\o}w}, E.,
  {Gonz{\'a}lez-Nuevo}, J., {G{\'o}rski}, K.~M., {Gratton}, S., {Gregorio}, A.,
  {Gruppuso}, A., {Gudmundsson}, J.~E., {Hamann}, J., {Hansen}, F.~K.,
  {Hanson}, D., {Harrison}, D.~L., {Helou}, G., {Henrot-Versill{\'e}}, S.,
  {Hern{\'a}ndez-Monteagudo}, C., {Herranz}, D., {Hildebrandt}, S.~R., {Hivon},
  E., {Hobson}, M., {Holmes}, W.~A., {Hornstrup}, A., {Hovest}, W., {Huang},
  Z., {Huffenberger}, K.~M., {Hurier}, G., {Jaffe}, A.~H., {Jaffe}, T.~R.,
  {Jones}, W.~C., {Juvela}, M., {Keih{\"a}nen}, E., {Keskitalo}, R., {Kisner},
  T.~S., {Kneissl}, R., {Knoche}, J., {Knox}, L., {Kunz}, M., {Kurki-Suonio},
  H., {Lagache}, G., {L{\"a}hteenm{\"a}ki}, A., {Lamarre}, J.~M., {Lasenby},
  A., {Lattanzi}, M., {Lawrence}, C.~R., {Leahy}, J.~P., {Leonardi}, R.,
  {Lesgourgues}, J., {Levrier}, F., {Lewis}, A., {Liguori}, M., {Lilje}, P.~B.,
  {Linden-V{\o}rnle}, M., {L{\'o}pez-Caniego}, M., {Lubin}, P.~M.,
  {Mac{\'\i}as-P{\'e}rez}, J.~F., {Maggio}, G., {Maino}, D., {Mandolesi}, N.,
  {Mangilli}, A., {Marchini}, A., {Maris}, M., {Martin}, P.~G., {Martinelli},
  M., {Mart{\'\i}nez-Gonz{\'a}lez}, E., {Masi}, S., {Matarrese}, S., {McGehee},
  P., {Meinhold}, P.~R., {Melchiorri}, A., {Melin}, J.~B., {Mendes}, L.,
  {Mennella}, A., {Migliaccio}, M., {Millea}, M., {Mitra}, S.,
  {Miville-Desch{\^e}nes}, M.~A., {Moneti}, A., {Montier}, L., {Morgante}, G.,
  {Mortlock}, D., {Moss}, A., {Munshi}, D., {Murphy}, J.~A., {Naselsky}, P.,
  {Nati}, F., {Natoli}, P., {Netterfield}, C.~B., {N{\o}rgaard-Nielsen}, H.~U.,
  {Noviello}, F., {Novikov}, D., {Novikov}, I., {Oxborrow}, C.~A., {Paci}, F.,
  {Pagano}, L., {Pajot}, F., {Paladini}, R., {Paoletti}, D., {Partridge}, B.,
  {Pasian}, F., {Patanchon}, G., {Pearson}, T.~J., {Perdereau}, O., {Perotto},
  L., {Perrotta}, F., {Pettorino}, V., {Piacentini}, F., {Piat}, M.,
  {Pierpaoli}, E., {Pietrobon}, D., {Plaszczynski}, S., {Pointecouteau}, E.,
  {Polenta}, G., {Popa}, L., {Pratt}, G.~W., {Pr{\'e}zeau}, G., {Prunet}, S.,
  {Puget}, J.~L., {Rachen}, J.~P., {Reach}, W.~T., {Rebolo}, R., {Reinecke},
  M., {Remazeilles}, M., {Renault}, C., {Renzi}, A., {Ristorcelli}, I.,
  {Rocha}, G., {Rosset}, C., {Rossetti}, M., {Roudier}, G., {Rouill{\'e}
  d'Orfeuil}, B., {Rowan-Robinson}, M., {Rubi{\~n}o-Mart{\'\i}n}, J.~A.,
  {Rusholme}, B., {Said}, N., {Salvatelli}, V., {Salvati}, L., {Sandri}, M.,
  {Santos}, D., {Savelainen}, M., {Savini}, G., {Scott}, D., {Seiffert}, M.~D.,
  {Serra}, P., {Shellard}, E.~P.~S., {Spencer}, L.~D., {Spinelli}, M.,
  {Stolyarov}, V., {Stompor}, R., {Sudiwala}, R., {Sunyaev}, R., {Sutton}, D.,
  {Suur-Uski}, A.~S., {Sygnet}, J.~F., {Tauber}, J.~A., {Terenzi}, L.,
  {Toffolatti}, L., {Tomasi}, M., {Tristram}, M., {Trombetti}, T., {Tucci}, M.,
  {Tuovinen}, J., {T{\"u}rler}, M., {Umana}, G., {Valenziano}, L., {Valiviita},
  J., {Van Tent}, F., {Vielva}, P., {Villa}, F., {Wade}, L.~A., {Wandelt},
  B.~D., {Wehus}, I.~K., {White}, M., {White}, S.~D.~M., {Wilkinson}, A.,
  {Yvon}, D., {Zacchei}, A., \& {Zonca}, A. 2016, \aap, 594, A13

\bibitem[{{Puchwein} {et~al.}(2018){Puchwein}, {Haardt}, {Haehnelt}, \&
  {Madau}}]{Puchwein:2018}
{Puchwein}, E., {Haardt}, F., {Haehnelt}, M.~G., \& {Madau}, P. 2018, ArXiv
  e-prints

\bibitem[{{Rahmati} {et~al.}(2013{\natexlab{a}}){Rahmati}, {Pawlik},
  {Rai{\v{c}}evi{\'c}}, \& {Schaye}}]{Rahmati2013MNRAS.430.2427R}
{Rahmati}, A., {Pawlik}, A.~H., {Rai{\v{c}}evi{\'c}}, M., \& {Schaye}, J.
  2013{\natexlab{a}}, \mnras, 430, 2427

\bibitem[{{Rahmati} {et~al.}(2013{\natexlab{b}}){Rahmati}, {Pawlik},
  {Rai{\v{c}}evi{\'c}}, \& {Schaye}}]{Rahmati:2013}
---. 2013{\natexlab{b}}, \mnras, 430, 2427

\bibitem[{{Rauch} {et~al.}(1997){Rauch}, {Miralda-Escud{\'e}}, {Sargent},
  {Barlow}, {Weinberg}, {Hernquist}, {Katz}, {Cen}, \& {Ostriker}}]{Rauch:1997}
{Rauch}, M., {Miralda-Escud{\'e}}, J., {Sargent}, W. L.~W., {Barlow}, T.~A.,
  {Weinberg}, D.~H., {Hernquist}, L., {Katz}, N., {Cen}, R., \& {Ostriker},
  J.~P. 1997, \apj, 489, 7

\bibitem[{{Schaye} {et~al.}(2015){Schaye}, {Crain}, {Bower}, {Furlong},
  {Schaller}, {Theuns}, {Dalla Vecchia}, {Frenk}, {McCarthy}, {Helly},
  {Jenkins}, {Rosas-Guevara}, {White}, {Baes}, {Booth}, {Camps}, {Navarro},
  {Qu}, {Rahmati}, {Sawala}, {Thomas}, \&
  {Trayford}}]{Schaye2015MNRAS.446..521S}
{Schaye}, J., {Crain}, R.~A., {Bower}, R.~G., {Furlong}, M., {Schaller}, M.,
  {Theuns}, T., {Dalla Vecchia}, C., {Frenk}, C.~S., {McCarthy}, I.~G.,
  {Helly}, J.~C., {Jenkins}, A., {Rosas-Guevara}, Y.~M., {White}, S. D.~M.,
  {Baes}, M., {Booth}, C.~M., {Camps}, P., {Navarro}, J.~F., {Qu}, Y.,
  {Rahmati}, A., {Sawala}, T., {Thomas}, P.~A., \& {Trayford}, J. 2015, \mnras,
  446, 521

\bibitem[{{Semenov} {et~al.}(2016){Semenov}, {Kravtsov}, \&
  {Gnedin}}]{2016ApJ...826..200S}
{Semenov}, V.~A., {Kravtsov}, A.~V., \& {Gnedin}, N.~Y. 2016, \apj, 826, 200

\bibitem[{{Shull} {et~al.}(2015){Shull}, {Moloney}, {Danforth}, \&
  {Tilton}}]{Shull:2015}
{Shull}, J.~M., {Moloney}, J., {Danforth}, C.~W., \& {Tilton}, E.~M. 2015,
  \apj, 811, 3

\bibitem[{{Sijacki} {et~al.}(2015){Sijacki}, {Vogelsberger}, {Genel},
  {Springel}, {Torrey}, {Snyder}, {Nelson}, \& {Hernquist}}]{Sijacki:2015}
{Sijacki}, D., {Vogelsberger}, M., {Genel}, S., {Springel}, V., {Torrey}, P.,
  {Snyder}, G.~F., {Nelson}, D., \& {Hernquist}, L. 2015, \mnras, 452, 575

\bibitem[{{Somerville} \& {Dav{\'e}}(2015)}]{Somerville2015ARA&A..53...51S}
{Somerville}, R.~S. \& {Dav{\'e}}, R. 2015, \araa, 53, 51

\bibitem[{{Springel}(2010)}]{Springel:2010}
{Springel}, V. 2010, MNRAS, 401, 791

\bibitem[{{Springel} {et~al.}(2005){Springel}, {Di Matteo}, \&
  {Hernquist}}]{Springel:2005}
{Springel}, V., {Di Matteo}, T., \& {Hernquist}, L. 2005, \mnras, 361, 776

\bibitem[{{Springel} \& {Hernquist}(2003)}]{Springel:2003}
{Springel}, V. \& {Hernquist}, L. 2003, \mnras, 339, 289

\bibitem[{{Springel} {et~al.}(2018){Springel}, {Pakmor}, {Pillepich},
  {Weinberger}, {Nelson}, {Hernquist}, {Vogelsberger}, {Genel}, {Torrey},
  {Marinacci}, \& {Naiman}}]{Springel:2018}
{Springel}, V., {Pakmor}, R., {Pillepich}, A., {Weinberger}, R., {Nelson}, D.,
  {Hernquist}, L., {Vogelsberger}, M., {Genel}, S., {Torrey}, P., {Marinacci},
  F., \& {Naiman}, J. 2018, \mnras, 475, 676

\bibitem[{{Tilton} {et~al.}(2012){Tilton}, {Danforth}, {Shull}, \&
  {Ross}}]{Tilton:2012}
{Tilton}, E.~M., {Danforth}, C.~W., {Shull}, J.~M., \& {Ross}, T.~L. 2012,
  \apj, 759, 112

\bibitem[{{Tonnesen} {et~al.}(2017){Tonnesen}, {Smith}, {Kollmeier}, \&
  {Cen}}]{Tonnesen2017}
{Tonnesen}, S., {Smith}, B.~D., {Kollmeier}, J.~A., \& {Cen}, R. 2017, \apj,
  845, 47

\bibitem[{{Tripp} {et~al.}(2008){Tripp}, {Sembach}, {Bowen}, {Savage},
  {Jenkins}, {Lehner}, \& {Richter}}]{Tripp:2008}
{Tripp}, T.~M., {Sembach}, K.~R., {Bowen}, D.~V., {Savage}, B.~D., {Jenkins},
  E.~B., {Lehner}, N., \& {Richter}, P. 2008, \apjs, 177, 39

\bibitem[{{Viel} {et~al.}(2017){Viel}, {Haehnelt}, {Bolton}, {Kim}, {Puchwein},
  {Nasir}, \& {Wakker}}]{Viel:2017}
{Viel}, M., {Haehnelt}, M.~G., {Bolton}, J.~S., {Kim}, T.-S., {Puchwein}, E.,
  {Nasir}, F., \& {Wakker}, B.~P. 2017, MNRAS, 467, L86

\bibitem[{{Villaescusa-Navarro} {et~al.}(2020){Villaescusa-Navarro},
  {Angl{\'e}s-Alc{\'a}zar}, {Genel}, {Spergel}, {Somerville}, {Dave},
  {Pillepich}, {Hernquist}, {Nelson}, {Torrey}, {Narayanan}, {Li}, {Philcox},
  {La Torre}, {Delgado}, {Ho}, {Hassan}, {Burkhart}, {Wadekar}, {Battaglia}, \&
  {Contardo}}]{camels:2020}
{Villaescusa-Navarro}, F., {Angl{\'e}s-Alc{\'a}zar}, D., {Genel}, S.,
  {Spergel}, D.~N., {Somerville}, R.~S., {Dave}, R., {Pillepich}, A.,
  {Hernquist}, L., {Nelson}, D., {Torrey}, P., {Narayanan}, D., {Li}, Y.,
  {Philcox}, O., {La Torre}, V., {Delgado}, A.~M., {Ho}, S., {Hassan}, S.,
  {Burkhart}, B., {Wadekar}, D., {Battaglia}, N., \& {Contardo}, G. 2020, arXiv
  e-prints, arXiv:2010.00619

\bibitem[{{Vogelsberger} {et~al.}(2013){Vogelsberger}, {Genel}, {Sijacki},
  {Torrey}, {Springel}, \& {Hernquist}}]{Vogelsberger:2013}
{Vogelsberger}, M., {Genel}, S., {Sijacki}, D., {Torrey}, P., {Springel}, V.,
  \& {Hernquist}, L. 2013, \mnras, 436, 3031

\bibitem[{{Vogelsberger} {et~al.}(2014{\natexlab{a}}){Vogelsberger}, {Genel},
  {Springel}, {Torrey}, {Sijacki}, {Xu}, {Snyder}, {Bird}, {Nelson}, \&
  {Hernquist}}]{Vogelsberger:2014a}
{Vogelsberger}, M., {Genel}, S., {Springel}, V., {Torrey}, P., {Sijacki}, D.,
  {Xu}, D., {Snyder}, G., {Bird}, S., {Nelson}, D., \& {Hernquist}, L.
  2014{\natexlab{a}}, \nat, 509, 177

\bibitem[{{Vogelsberger} {et~al.}(2014{\natexlab{b}}){Vogelsberger}, {Genel},
  {Springel}, {Torrey}, {Sijacki}, {Xu}, {Snyder}, {Bird}, {Nelson}, \&
  {Hernquist}}]{Vogelsberger:2014}
---. 2014{\natexlab{b}}, Nature, 509, 177

\bibitem[{{Vogelsberger} {et~al.}(2014{\natexlab{c}}){Vogelsberger}, {Genel},
  {Springel}, {Torrey}, {Sijacki}, {Xu}, {Snyder}, {Nelson}, \&
  {Hernquist}}]{Vogelsberger:2014b}
{Vogelsberger}, M., {Genel}, S., {Springel}, V., {Torrey}, P., {Sijacki}, D.,
  {Xu}, D., {Snyder}, G., {Nelson}, D., \& {Hernquist}, L. 2014{\natexlab{c}},
  \mnras, 444, 1518

\bibitem[{{Vogelsberger} {et~al.}(2012){Vogelsberger}, {Sijacki},
  {Kere{\v{s}}}, {Springel}, \& {Hernquist}}]{Vogelsberger:2012}
{Vogelsberger}, M., {Sijacki}, D., {Kere{\v{s}}}, D., {Springel}, V., \&
  {Hernquist}, L. 2012, \mnras, 425, 3024

\bibitem[{{Walther} {et~al.}(2019{\natexlab{a}}){Walther}, {O{\~n}orbe},
  {Hennawi}, \& {Luki{\'c}}}]{Walther2019ApJ...872...13W}
{Walther}, M., {O{\~n}orbe}, J., {Hennawi}, J.~F., \& {Luki{\'c}}, Z.
  2019{\natexlab{a}}, \apj, 872, 13

\bibitem[{{Walther} {et~al.}(2019{\natexlab{b}}){Walther}, {O{\~n}orbe},
  {Hennawi}, \& {Luki{\'c}}}]{2019ApJ...872...13W}
---. 2019{\natexlab{b}}, \apj, 872, 13

\bibitem[{{Weinberger} {et~al.}(2017){Weinberger}, {Springel}, {Hernquist},
  {Pillepich}, {Marinacci}, {Pakmor}, {Nelson}, {Genel}, {Vogelsberger},
  {Naiman}, \& {Torrey}}]{Weinberger:2017}
{Weinberger}, R., {Springel}, V., {Hernquist}, L., {Pillepich}, A.,
  {Marinacci}, F., {Pakmor}, R., {Nelson}, D., {Genel}, S., {Vogelsberger}, M.,
  {Naiman}, J., \& {Torrey}, P. 2017, MNRAS, 465, 3291

\bibitem[{{Weinberger} {et~al.}(2020){Weinberger}, {Springel}, \&
  {Pakmor}}]{Weinberger2020ApJS..248...32W}
{Weinberger}, R., {Springel}, V., \& {Pakmor}, R. 2020, \apjs, 248, 32

\bibitem[{{Werk} {et~al.}(2016){Werk}, {Prochaska}, {Cantalupo}, {Fox},
  {Oppenheimer}, {Tumlinson}, {Tripp}, {Lehner}, \&
  {McQuinn}}]{werk2016ApJ...833...54W}
{Werk}, J.~K., {Prochaska}, J.~X., {Cantalupo}, S., {Fox}, A.~J.,
  {Oppenheimer}, B., {Tumlinson}, J., {Tripp}, T.~M., {Lehner}, N., \&
  {McQuinn}, M. 2016, \apj, 833, 54

\bibitem[{{Weymann} {et~al.}(1998){Weymann}, {Jannuzi}, {Lu}, {Bahcall},
  {Bergeron}, {Boksenberg}, {Hartig}, {Kirhakos}, {Sargent}, {Savage},
  {Schneider}, {Turnshek}, \& {Wolfe}}]{Weymann:1998}
{Weymann}, R.~J., {Jannuzi}, B.~T., {Lu}, L., {Bahcall}, J.~N., {Bergeron}, J.,
  {Boksenberg}, A., {Hartig}, G.~F., {Kirhakos}, S., {Sargent}, W.~L.~W.,
  {Savage}, B.~D., {Schneider}, D.~P., {Turnshek}, D.~A., \& {Wolfe}, A.~M.
  1998, \apj, 506, 1

\bibitem[{{Zaldarriaga} {et~al.}(2001){Zaldarriaga}, {Hui}, \&
  {Tegmark}}]{2001ApJ...557..519Z}
{Zaldarriaga}, M., {Hui}, L., \& {Tegmark}, M. 2001, \apj, 557, 519

\bibitem[{{Zhang} {et~al.}(1995){Zhang}, {Anninos}, \& {Norman}}]{Zhang:1995}
{Zhang}, Y., {Anninos}, P., \& {Norman}, M.~L. 1995, \apj, 453, L57

\bibitem[{{Zinger} {et~al.}(2020){Zinger}, {Pillepich}, {Nelson}, {Weinberger},
  {Pakmor}, {Springel}, {Hernquist}, {Marinacci}, \&
  {Vogelsberger}}]{Zinger2020MNRAS.499..768Z}
{Zinger}, E., {Pillepich}, A., {Nelson}, D., {Weinberger}, R., {Pakmor}, R.,
  {Springel}, V., {Hernquist}, L., {Marinacci}, F., \& {Vogelsberger}, M. 2020,
  \mnras, 499, 768

\end{thebibliography}
 
\end{document}